\documentclass[12pt,twoside]{article}  
\usepackage[super,sort,comma]{natbib}
\usepackage{silence}
\usepackage{xcolor}
\usepackage{fancyhdr}	
 
\newcommand{\captionv}[3]{\begin{center}\parbox{#1cm}{\caption[#2]{{\sf #3}}}\end{center}}
\usepackage[section]{placeins}   
\usepackage{graphicx}
\makeatletter \renewcommand\@biblabel[1]{$^{#1}$} \makeatother
\newcommand{\cen}[1]{\begin{center} #1 \end{center}}

\newenvironment{packed_item}{
\begin{itemize}
\setlength{\itemsep}{1pt}
\setlength{\parskip}{0pt}
\setlength{\parsep}{0pt}
}{\end{itemize}}
   
\usepackage[pagewise,mathlines,edtable]{lineno}
\usepackage{hyperref}
\hypersetup{ colorlinks,citecolor=blue,filecolor=blue,linkcolor=blue,urlcolor=blue}
\usepackage{xcolor}
\definecolor{gray}{rgb}{0.6,0.6,0.6}
\definecolor{red}{rgb}{0.85,0,0}
\definecolor{green}{rgb}{0,0.85,0}
\definecolor{blue}{rgb}{0,0,0.85}
\definecolor{beige}{rgb}{0.92,0.87,0.78}
\usepackage[all]{hypcap}
\usepackage{threeparttable}
\usepackage{amsmath}
\usepackage{tikz}
\usepackage{pgfplots}
\usetikzlibrary{shadows,arrows}
\usepackage{subcaption}
\pgfplotsset{compat = 1.14}
\usepackage{makecell}
\usepackage{multirow}
\usepackage{comment}
\usepackage[all]{hypcap} 
\tikzstyle{block} = [draw,fill=blue!10, inner sep=0cm, text centered, minimum height=1.5em, rounded corners, very thick, minimum width=5.5em]
\tikzstyle{box} = [draw, inner sep=0cm, minimum height=19em, rounded corners, very thick, minimum width=17.5em]
\usetikzlibrary{positioning,arrows.meta}
\usetikzlibrary{shapes,arrows,chains}
\usetikzlibrary[calc]
\tikzstyle{wh} = [text width=1.5em, text centered]
\tikzstyle{cr} = [draw, circle,fill=blue!10, inner sep=0cm,text width=1em, text centered]
\tikzstyle{blak} = [inner sep=0cm]
\usetikzlibrary{positioning,arrows.meta}
\newcommand\ImageNode[3][]{\node[draw=black,line width=3pt,#1, inner sep=0cm] (#2) {\includegraphics[scale=0.25]{#3}};}

\usepackage{amsmath}
\usepackage{pifont}
\newcommand{\cmark}{\text{\ding{51}}}
\newcommand{\xmark}{\text{\ding{55}}}
\usepackage{color,soul, colortbl}
\definecolor{reviewer1}{rgb}{1,1,1}
\definecolor{reviewer2}{rgb}{1,1,1}
\newcommand{\hlfr}[1]{{\sethlcolor{reviewer1}\hl{#1}}}

\begin{document}
\cen{\sf {\Large {\bfseries MRQy: An Open-Source Tool for Quality Control\\ of MR Imaging Data}\\  
\vspace*{10mm}
Amir Reza Sadri$^{1}$, Andrew Janowczyk$^{1}$, Ren Zhou$^{1}$, Ruchika Verma$^{1}$, \\ 
Niha Beig$^{1}$, Jacob Antunes$^{1}$, Anant Madabhushi$^{1,2}$, Pallavi Tiwari$^{1}$, \\ \vspace*{3mm}
and Satish E Viswanath$^{1}$} \\ \vspace*{3mm}
{\small{$^{1}$ Department of Biomedical Engineering, Case Western Reserve University, Cleveland, OH 44106, USA \\
$^{2}$ Louis Stokes Cleveland VA Medical Center, Cleveland, OH 44106, USA}}
\vspace{5mm}\\
Version typeset \today\\
Corresponding author: sev21@case.edu \\
}
\pagenumbering{roman}
\setcounter{page}{1}
\pagestyle{plain}

\begin{abstract}
\noindent 
{\bf Purpose:} We sought to develop a quantitative tool to quickly determine relative differences in MRI volumes both within and between large MR imaging cohorts (such as available in The Cancer Imaging Archive (TCIA)), in order to help determine the generalizability of radiomics and machine learning schemes to unseen datasets. 
The tool is intended to help quantify presence of (a) site- or scanner-specific variations in image resolution, field-of-view, or image contrast, or (b) imaging artifacts such as noise, motion, inhomogeneity, ringing, or aliasing; which can adversely affect relative image quality between data cohorts.
\\
{\bf Methods:} We present MRQy, a new open-source quality control tool to (a) interrogate MRI cohorts for site- or equipment-based differences, and (b) quantify the impact of MRI artifacts on relative image quality; to help determine how to correct for these variations prior to model development. MRQy extracts a series of quality measures (e.g. noise ratios, variation metrics, entropy and energy criteria) and MR image metadata (e.g. voxel resolution, image dimensions) for subsequent interrogation via a specialized HTML5 based front-end designed for real-time filtering and trend visualization. \\
{\bf Results:} MRQy was used to evaluate (a) n=133 brain MRIs from TCIA (7 sites), and (b) n=104 rectal MRIs (3 local sites). MRQy measures revealed significant site-specific variations in both cohorts, indicating potential batch effects. 
Marked differences in specific MRQy measures were also able to identify outlier MRI datasets that needed to be corrected for common MR imaging artifacts. \\
{\bf Conclusions:} MRQy is designed to be a standalone, unsupervised tool that can be efficiently run on a standard desktop computer. It has been made freely accessible at \url{http://github.com/ccipd/MRQy} for wider community use and feedback. \\
\end{abstract}

\newpage    
\tableofcontents
\newpage
\setlength{\baselineskip}{0.7cm}      		
\pagenumbering{arabic}
\setcounter{page}{1}
\pagestyle{fancy}

\section{Introduction}
\begin{figure}[t!]
\begin{center}
\resizebox {1\textwidth}{!}{
\fbox{\includegraphics[width = \textwidth, height = 9cm]{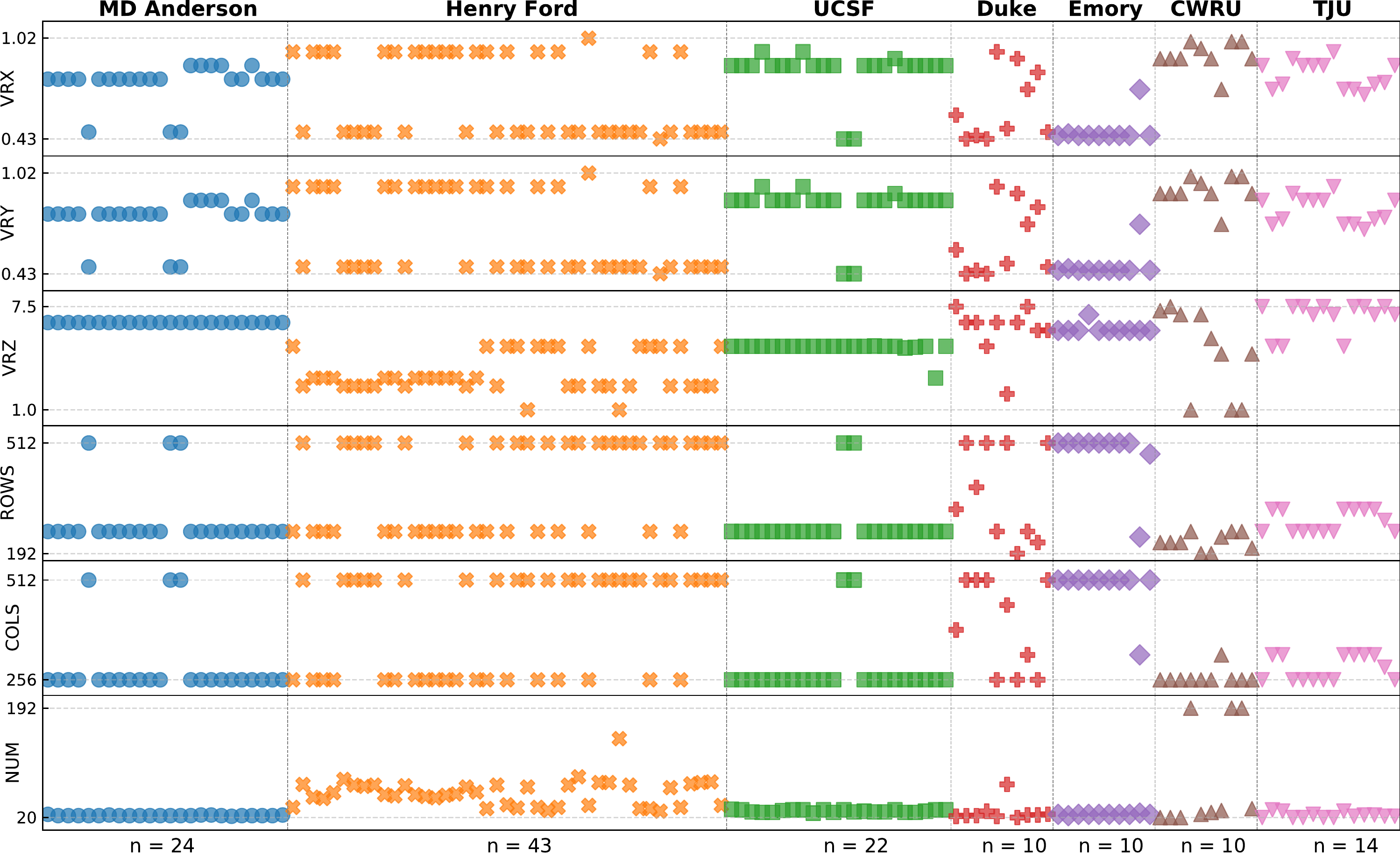}}%
}
\captionv{16}{Short title - can be blank}
{Variations in MRI scan metadata illustrating batch effects within the TCGA-GBM cohort across 133 {\sethlcolor{reviewer1}\hl{post-contrast T1-weighted MRI datasets}} curated from 7 different sites (in different colors) where each point corresponds to a unique subject. $VRX$, $VRY$, and $VRZ$ correspond to voxel resolutions {\sethlcolor{reviewer1}\hl{in-plane (x, y) and through-plane (z)}} respectively. $ROWS$ and $COLS$ represent in-plane scan size and the $NUM$ is the number of images in each volume. All information was directly extracted from DICOM metadata of each MRI dataset, as downloaded from TCIA.   
\label{fig_fore_back}}
\end{center}
\end{figure}

The development of public repositories such as The Cancer Imaging Archive (TCIA)~\cite{clark2013cancer, prior2017public} have enabled significant advances in machine and deep learning approaches via radiographic imaging for a variety of oncological applications~\cite{kalpathy2014quantitative,zanfardino2019tcga}.
With over 2500 MR imaging scans for 26 unique anatomic sites obtained from different centers and using a variety of scanner equipment, a key aspect to utilizing TCIA cohorts for reliable model development and optimization of computational imaging tools~\cite{prior2020open} is to curate datasets with minimal to no artifacts; implying they are relatively homogeneous in appearance~\cite{basu2019call}. Quantitative assessment of variations or artifacts known to exist in MRI data~\cite{schlett2016quantitative} can also provide \textit{a priori} cues regarding the generalizability of computational imaging models to unseen datasets, by helping identify:
\begin{packed_item}
\item \textit{Site- and scanner-specific variations} which can cause poor reproducibility of machine learning models between cohorts and sites\cite{glocker2019machine, wachinger2019quantifying}. Beyond just differences in image acquisition parameters such as echo and repetition times, there may also be systematic occurrences of technical variations (e.g. voxel
resolution or fields-of-view) in subsets of a cohort i.e. \textit{batch effects}~\cite{leek2010tackling}. These are exemplified by distinctive variations in image and voxel dimensions both within as well as across 7 distinct sites within The Cancer Genome Atlas Glioblastoma Multiforme (TCGA-GBM) collection~\cite{tcia} in Figure \ref{fig_fore_back}. 
\item \textit{Presence of imaging artifacts} which adversely affect the relative quality of clinical MR imaging scans within a cohort and significantly degrade model performance~\cite{zwanenburg2020image}. Issues such as magnetic field inhomogeneity, aliasing, motion, ringing, or noise~\cite{erasmus2004short,Bushberg} can cause individual datasets to deviate significantly from the remainder of the cohort; and need to be identified and corrected prior to analysis~\cite{um2019impact,schwier2019repeatability}.
\end{packed_item}
\noindent
Evaluating variations and relative image quality between cohorts is thus critical to determine whether a radiomics or deep learning model that was trained on one cohort will perform reproducibly on a different cohort.
{\sethlcolor{reviewer2}\hl{Identification of artifacts in individual MRI datasets (known as image \textit{quality assessment})  may be performed by experts, however, visual inspection is known to lack sensitivity to subtle variations between MR images{~\cite{gardner1995detection}}. 
Subjective quality ratings are also not precise enough to rigorously curate MRI cohorts{~\cite{esteban2017mriqc}} due to inter-rater variability. Additionally, it may be infeasible to laboriously and manually assess the quality of individual imaging datasets in large public repositories such as TCIA.
This underscores the need for automated \textit{quality control (QC)} tools for MRI scans, defined in this context as quantitative approaches that can be used to identify ranges of acceptability for MR imaging datasets{~\cite{delis2017moving}}. Importantly, QC allows a user to quickly identify when a quality measurement falls outside a user-specified range or tolerance, so that an appropriate corrective action can be taken (for specific datasets or the cohort as a whole){\cite{Silvia2008}}.}} Such QC tools also need to work with minimal to no user intervention while enabling reliable curation of artifact-free and congruent data cohorts for further computational analysis and model development.

\setlength{\fboxsep}{10pt}
\setlength{\fboxrule}{1.5pt}

Efforts towards automated quality control of MRI datasets has led to development of different Image Quality Metrics (IQMs)~\cite{Ieremeiev2018, woodard2006no} as well as a Quality Assessment Protocol (QAP)~\cite{VanEssen2012}. However, most studies evaluating IQMs (e.g. noise ratios, energy ratios, entropy values) have focused on supervised prediction of whether a brain MRI should be {\textit{accepted}} or {\textit{excluded}}, as defined by experts~\cite{woodard2006no,Mortamet2009,Pizarro2016}.
IQMs have also been implemented in tools such as MRIQC~\cite{esteban2017mriqc} which is a supervised quality classification tool for predicting expert quality ratings for brain MRIs. More recently, MRIQC was adapted to use a web-application program interface to facilitate efficient interactions with expert quality ratings and quality annotations~\cite{esteban2019crowdsourced}.
Other quality control and prediction tools for brain MRIs include the FreeSurfer-specific Qoala-T~\cite{KLAPWIJK2019116} and LAB–QA2GO~\cite{vogelbacher2019lab}; where the latter outputs an HTML report of image quality measures.
{\sethlcolor{reviewer2}\hl{However, as summarized in Table {\ref{comp}}, none of these tools{~\cite{esteban2017mriqc,esteban2019crowdsourced,KLAPWIJK2019116,vogelbacher2019lab,KESHAVAN2018365,pradeep_reddy_raamana_2018_1211365}} are readily generalizable to MRIs of body regions other than the brain, or provide an interactive front-end that can be used to easily interrogate quality issues and batch effects that may be present in large-scale imaging cohorts.}}

\setlength{\tabcolsep}{2pt}
\def\arraystretch{1.02}
\begin{table}[t!]
\scriptsize
\begin{center}
\caption{\protect\sethlcolor{reviewer2}\hl{Feature comparison of open-source MRI quality control tools}}
\label{comp}
\resizebox {1\textwidth}{!}{
\begin{threeparttable}
\begin{tabular}{l|ccccccc}
\Xhline{2\arrayrulewidth}
\rowcolor{reviewer2}\hspace{2cm}{\scriptsize{\textbf{Attribute}}} & {\scriptsize{MRIQC}}\cite{esteban2017mriqc} &{\scriptsize{Qoala-T}}\cite{KLAPWIJK2019116}&{\scriptsize{LAB–QA2GO}}\cite{vogelbacher2019lab}&{\scriptsize{MRIQC-WebAPI}}\cite{esteban2019crowdsourced}&
{\scriptsize{Mindcontrol}}\cite{KESHAVAN2018365}&{\scriptsize{VisualQC}}\cite{pradeep_reddy_raamana_2018_1211365}&{\scriptsize{MRQy}}\\
\Xhline{2\arrayrulewidth}
\rowcolor{reviewer2} Body part& brain & brain & brain & brain & brain & brain&all\\
\rowcolor{reviewer2} Supporting raw data ({\textit{.dcm},{\textit{.nii}}})& \cmark & \cmark & \cmark &\cmark&\cmark&\cmark& \cmark\\
\rowcolor{reviewer2} Supporting processed data ({\textit{.mha}})& \xmark & \xmark & \xmark &\xmark&\xmark&\cmark& \cmark \\
\rowcolor{reviewer2}Supporting  modular plugins & \xmark & \xmark & \xmark &\xmark &\cmark&\cmark&\cmark\\
\rowcolor{reviewer2} Supporting phantom data& \xmark & \xmark & \cmark &\xmark &\xmark&\xmark&\cmark\\
\rowcolor{reviewer2} Interactive user interface & \cmark & \xmark & \xmark &\cmark&\cmark&\cmark& \cmark\\
\rowcolor{reviewer2} Batch effects identification& \cmark & \xmark & \xmark &\cmark&\xmark&\cmark& \cmark\\
\rowcolor{reviewer2} Measurements& 14 & 185 & - & 14 & 3+9 & - & 13+10\\
\Xhline{2\arrayrulewidth}
\end{tabular}
\begin{center}
\end{center}
\end{threeparttable}
}
\end{center}
\end{table}

In this work, we present a technical overview of MRQy, a new open-source quality control tool for MR imaging data. 
MRQy builds on the HistoQC Python framework~\cite{janowczyk2019histoqc}  
and has been specialized for analyzing large-scale MRI cohorts through the following modules: (i) automatic foreground detection for any MR image from any body region, from which it will (ii) extract a series of imaging-specific metadata and quality measures\cite{VanEssen2012} generalized to work with any structural MR sequence, in order to (iii) compute representations that capture relevant MR image quality trends in a data cohort.
{\sethlcolor{reviewer2}\hl{These are presented within a specialized HTML5-based front-end which can be used to: (a) interrogate trends in per-site and per-scanner variations in a multi-site setting, (b) identify which specific image artifacts are present in which MRI scans in a data cohort, and (c) curate together cohorts of MRI datasets which are consistent and of sufficient image quality for computational model development.}}
{\sethlcolor{reviewer1}\hl{We will demonstrate the usage of MRQy via a representative publicly available brain MRI cohort from TCIA as well as an in-house multi-site rectal MRI cohort.}}
MRQy works in an unsupervised standalone setting, runs efficiently on a standard computer, and has a modular design to allow for easy incorporation of additional algorithms and metrics as plugins in the future. 

\section{Materials and Methods}
\begin{figure}[t!]
\begin{center}
\resizebox {1\textwidth}{!}{
\begin{tikzpicture}[node distance=1cm,>={Triangle[angle=60:1pt 2]},shorten >= 0pt,shorten <= 0pt,arrow/.style={-latex',blue!40,line width=2pt}]
\node[block] (I) {\footnotesize{Dataset}};
\node[block, right=of I] (FB) {\footnotesize{FG Detection}};
\node[block, right=of FB] (M) {\footnotesize{Measures}};
\node[block, below=of M] (T) {\footnotesize{Metadata}};
\node[block, below=of T] (Th) {\footnotesize{Thumbnails}};
\node[block, right=of M] (TS) {\footnotesize{t-SNE}};
\node[block, right=of T] (U) {\footnotesize{UMAP}};
\node[box, right=of TS] (B) {};
\draw[arrow](I) -- (FB);
\draw[arrow](FB) -- (M);
\draw[arrow](M) -- (TS);
\draw[arrow](T) -| ($(M) + (1.4,0)$);
\draw[arrow] ($(M) + (1.7,0)$) |- (U);
\draw[arrow](TS) -- (B);
\draw[arrow](U) -| ($(TS) + (1.4,0)$);
\draw[arrow](Th) -- ($(Th) + (5.5,0)$);
\draw[arrow]($(I) - (-0.15,0.33)$) |- (T);
\draw[arrow]($(I) - (0.15,0.33)$) |- (Th);
\node[draw,fill=blue!10, inner sep=0cm, text centered, minimum height=1.5em, rounded corners, very thick, minimum width=6em,align=center] at (15.75,3.55)
{\footnotesize{Tables}};
\node(TT) at (15.75,2.45) {\includegraphics[width=7cm, height = 1.4cm]{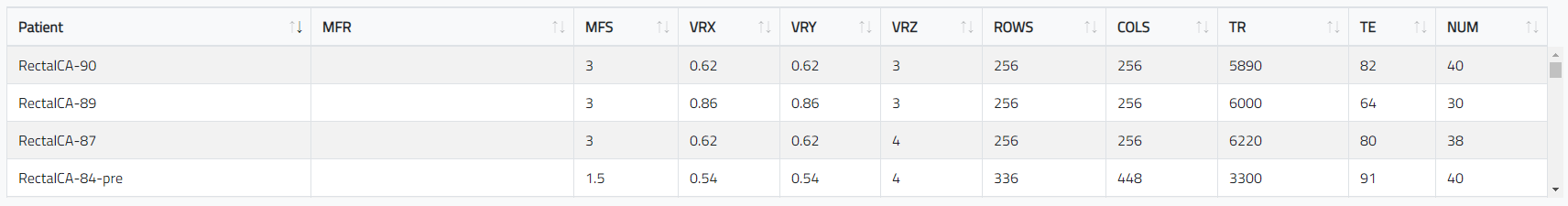}};
\node[draw,fill=blue!10, inner sep=0cm, text centered, minimum height=1.5em, rounded corners, very thick, minimum width=5.5em,align=center]at (15.75,1.35) {\footnotesize{Charts}};
\node (cp) at (15.75,0.1) {\includegraphics[width=7cm, height = 1.5cm]{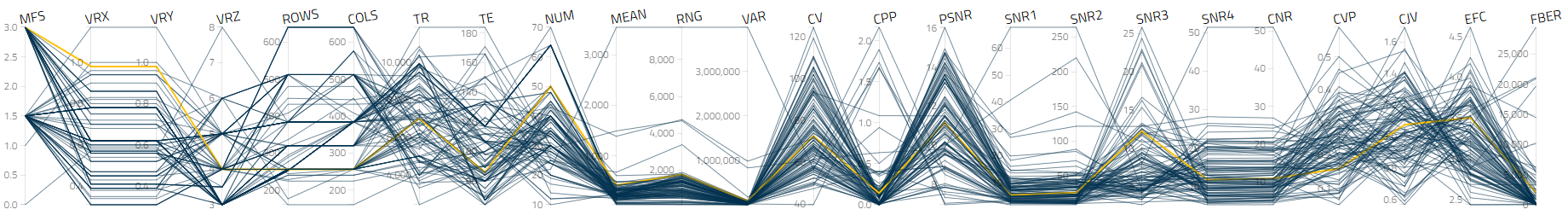}};
\node(bp) at (15.75,-1.3) {\includegraphics[width=7cm, height = 1.4cm]{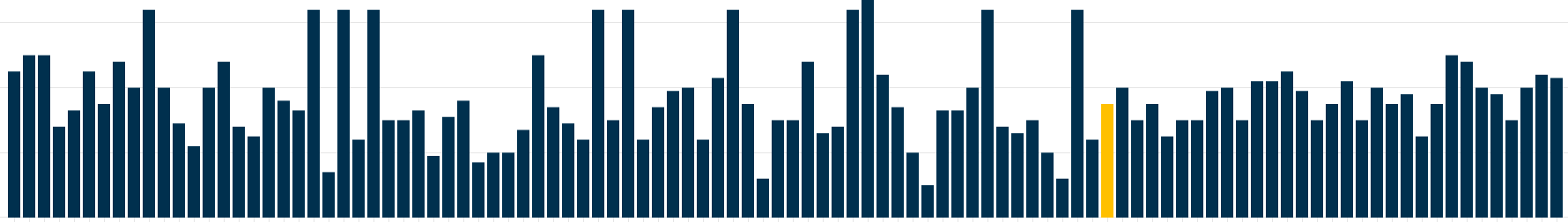}};
\node[draw,fill=blue!10, inner sep=0cm, text centered, minimum height=1.5em, rounded corners, very thick, minimum width=5.5em,align=center] at (15.75,-2.45) {\footnotesize{Plots}};
\node (bp) at (15.75,-3.4) {\includegraphics[width=5.5cm, height = 0.9cm]{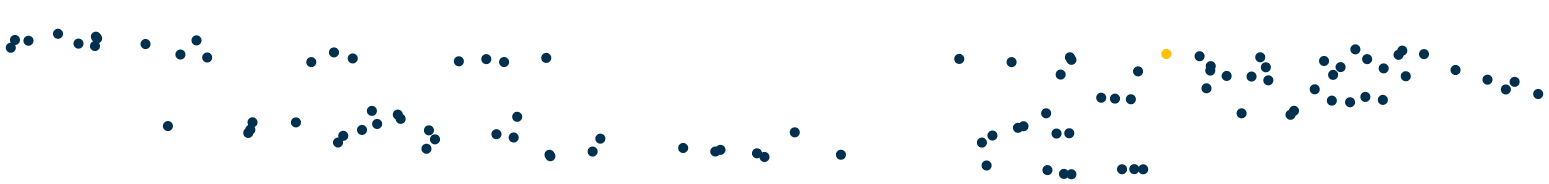}};
\draw[draw, dashed, color = gray] ($(M) + (-1.55,0.7)$) -- ($(T) + (-1.55,-0.7)$)  -- ($(T) + (1.55,-0.7)$) -- ($(M) + (1.55,0.7)$) -- cycle;
\draw[draw, dashed, color = black!60] ($(FB) + (-1.7,1.2)$) -- ($(FB) + (-1.7,-4.025)$)  -- ($(Th) + (2,-0.7)$) -- ($(M) + (2,1.2)$) -- cycle;
\node at ($(I) + (0,0.75)$) {\scriptsize{\bf{Input}}};
\node at ($(M) + (0,1)$) {\scriptsize{Information Extraction}};
\node at ($(M) + (-1.6,1.5)$) {\scriptsize{\bf{Backend Processing}}};
\node at ($(B) + (-0.1,4.2)$) {\scriptsize{\bf{Front-end Visualization}}};
\node[rotate=90] at ($(B) + (-3.87,0.8)$) {\scriptsize{\it{results.tsv}}};
\node[rotate=0] at ($(Th) + (4.2,0.3)$) {\scriptsize{\it{thumbnails.png}}};
\end{tikzpicture}
}
\captionv{15}{Short title - can be blank}
{Schematic for overall MRQy workflow and major components.
\label{fig_workflow} 
}  
\end{center}
\end{figure}

The major components of the MRQy tool are illustrated in Figure \ref{fig_workflow}, which can be sub-divided into three specific modules: Input, Backend Processing, and Front-end Visualization. 

\subsection{Input}
To ensure MRQy supported popular file formats for storing radiographic data (e.g.  \textit{.dcm}, \textit{.nii}, \textit{.mha}, and \textit{.ima}), the packages \verb|medpy| and \verb|pydicom| libraries~\cite{pypi} were used. MRQy iteratively parses either a single directory input (containing files) or resources through a directory of directories, in order to read in the image volume as well as parse image metadata.

\subsection{Backend Processing}

\subsubsection{Foreground Detection}

\begin{figure}[bth!]
\begin{center}
\resizebox {1\textwidth}{!}{
\begin{tikzpicture}[
  node distance=1.5cm,
  >={Triangle[angle=60:1pt 2]},
  shorten >= 0pt,
  shorten <= 0pt,
  arrow/.style={
    -latex',
    blue!40,
    line width=2pt
  }
]
\ImageNode[label={}]{I}{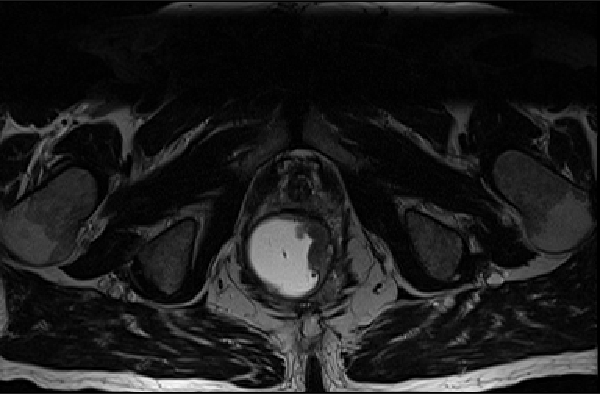}
\node [cr, right=of I] (M1) {${\large{\times}}$};
\ImageNode[label={90:\footnotesize{estimated background}},right=of M1]{S1}{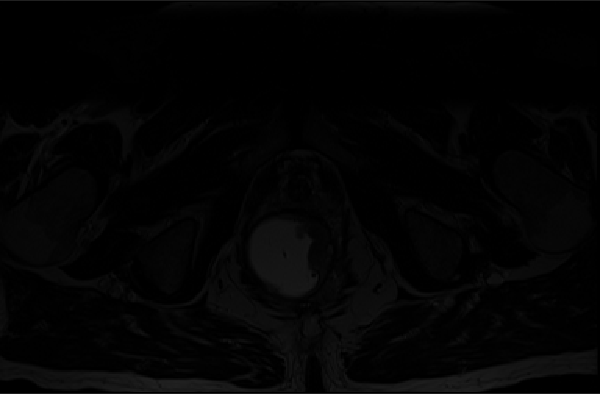}
\ImageNode[below=of I]{H}{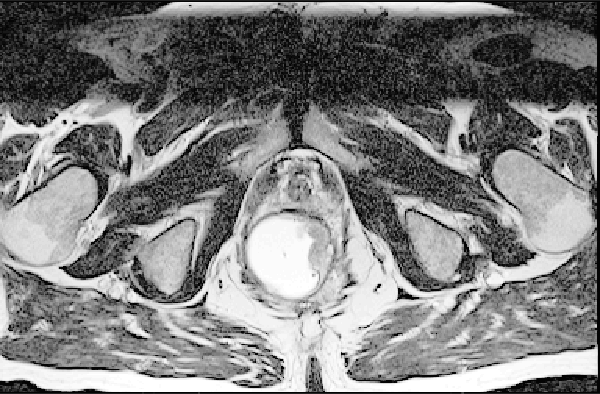}
\node [cr, right=of H] (M2) {${\large{\times}}$};
\ImageNode[label={-90:\footnotesize{estimated foreground}},right=of M2]{S2}{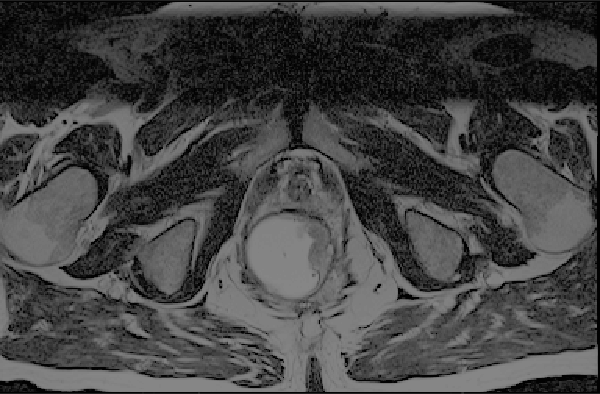}
\node[cr, below=of S1, yshift=1cm] (SU) {${\large{+}}$};
\ImageNode[label={90:\footnotesize{combined volume}},right=of S2, xshift= -1.2cm, yshift=1.65cm]{S}{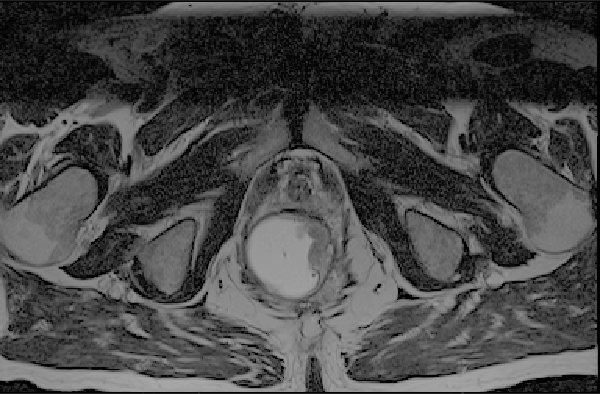}
\ImageNode[label={90:\footnotesize{overlay mask}},right=of S, yshift= 0cm, xshift= -0.5]{OL}{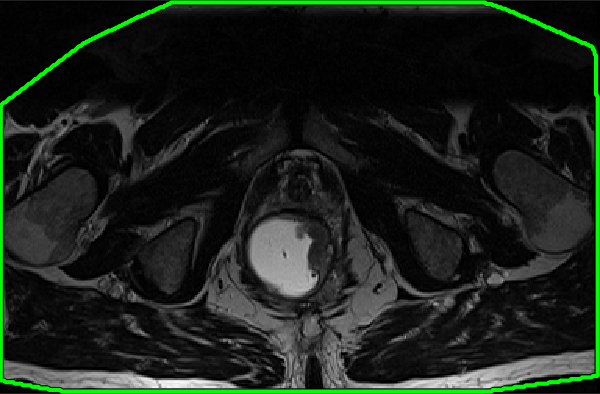}
\draw[arrow](I) -- (H);
\draw[arrow](I) -- (M1);
\draw[arrow](H) -- (M2);
\draw[arrow]($(I) + (0,0.9)$) -- ($(I) + (0,1.2)$)  -| (M1);
\draw[arrow]($(H) + (0,-0.9)$) -- ($(H) + (0,-1.2)$)  -| (M2);
\draw[arrow](M1) -- (S1);
\draw[arrow](M2) -- (S2);
\draw[arrow](S1) -- (SU);
\draw[arrow](S2) -- (SU);
\draw[arrow](SU) -- (S);
\draw[arrow](S) -- (OL);
\node at (-0.3,-1.6) {\it{\footnotesize{he}}};
\node at (1.3,-1.2) {\footnotesize{original volume}};
\node at (1.8,-2.2) {\footnotesize{histogram equalized}};
\node at (1.8,1.4) {\it{\footnotesize{mean(otsu)}}};
\node at (3.62,0.58) {\footnotesize{$w_1$}};
\node at (3.62,-4) {\footnotesize{$w_2$}};
\node at (1.8,-4.8) {\it{\footnotesize{mean(otsu)}}};
\node at (11.1,-1.2) {\it{\footnotesize{ch(otsu)}}};
\end{tikzpicture}
}
\captionv{15}{Short title - can be blank}
{Schematic of the foreground detection algorithm.
\label{fb_alg} 
}  
\end{center}
\end{figure}

As the MRI volume input to MRQy could be acquired from any body region, it is crucial to first identify the primary area within the volume from which quality measures are calculated. An in-house image processing algorithm was developed to efficiently and automatically extract and separate the background (outside the body) from the foreground (primary region of interest), as illustrated in Figure \ref{fb_alg}.

Otsu thresholding is typically proposed as an efficient approach to quickly identify both these regions in an MRI volume, but common shadowing/shading artifacts that are present in MRI volumes are known to significantly affect its performance{\sethlcolor{reviewer1}\hl{{\cite{FENG2017186}}}}.
By contrast, histogram equalization could help articulate details in the MRI volume even with shading or shadow artifacts, but is known to also intensify the appearance of noise{\sethlcolor{reviewer1}\hl{{\cite{TORRENTSBARRENA2019263}}}}.
In order to take advantage of both methods simultaneously, a weighted combination of the original MRI volume and its histogram equalized version were used as an input to Otsu thresholding. The weights were defined automatically to ensure a completely unsupervised method. A convex hull operation was then used to enclose the foreground based on the mask output from Otsu thresholding. \hlfr{While this algorithm was designed primarily for scans with one primary foreground region, it has been generalized to be able to detect and return multiple individual foreground objects that may be present in an MR image (e.g. axial slices over both legs{~\cite{7994618}}). The Supplementary Materials section has additional details on settings that can be specified by the end-user in this regard.}

\begin{figure}[t!]
\begin{center}
\resizebox {1\textwidth}{!}{
\includegraphics[width=\textwidth]{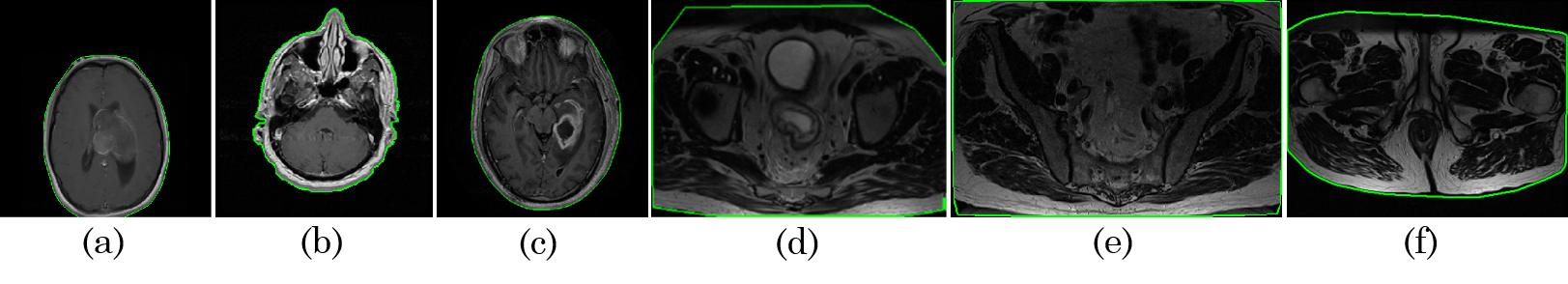}
}
\captionv{15}{}
{Representative foreground detection results with the foreground outlined in green for representative (a), (b), (c) brain MRI sections, (d), (e), (f) rectal MRI sections. Region outside the green outline is considered background. Note accurate delineation of outer boundary despite visually apparent shadowing and noise artifacts on all MRIs, regardless of body region being imaged.
\label{fb_examp}}  
\end{center}
\end{figure}

These steps were implemented using the \verb|scikit-image| python package~\cite{scikit-image}. Representative results of foreground detection are illustrated in Figure \ref{fb_examp} for MRIs of the brain and the rectum to demonstrate the generalizability of the algorithm across different body regions, \hlfr{including in MR images with minimal background (i.e. ``zoomed" acquisitions or those with sufficient phase oversampling, see Figure {\ref{fb_examp}}(e))}.
\par 

\captionsetup[sub]{font=small,labelfont=small}

\subsubsection{Information Extraction}
Two major types of information were extracted from each MRI volume, broadly categorized as {\it{Metadata}} and {\it{Measurements}} (summarized in Table \ref{table}). These were saved into a tab-separated file for further analysis.
\begin{packed_item}
\item
{\it\underline{Metadata:}}
This information that was directly extracted from file headers, such as voxel resolution or MRI volume dimensions (summarized in Table \ref{table}, rows 1-10). 
\item
{\it\underline{Measurements:}}
These were selected from a survey of the medical imaging literature~\cite{chang2015reference,dietrich2007measurement,esteban2017mriqc,VanEssen2012} for detecting specific artifacts in MRI scans. 
This list included statistical measures (e.g. range, variance, \%CV) as well as second-order statistics and filter-based measures (e.g. contrast per pixel (CPP), entropy focus criterion (EFC), signal-to-noise ratios of different regions). Table \ref{table} (rows 13-23) summarizes the measures extracted by MRQy, their mathematical formulation, and what type of artifact they were each intended to quantify.
\end{packed_item}

\setlength{\tabcolsep}{2pt}
\def\arraystretch{1.02}
\begin{table}[ht!]
\scriptsize
\begin{center}
\caption{Summary table of metadata and quality measures extracted within MRQy}
\label{table}
\resizebox {1\textwidth}{!}{
\begin{threeparttable}
\begin{tabular}{c|cp{13.5cm}}
\Xhline{2\arrayrulewidth}
{\scriptsize{\textbf{Type}}} & {\scriptsize{\textbf{Metric}}} & {\scriptsize{\textbf{Description}}\tnote{*}} \\
\Xhline{2\arrayrulewidth}
\multirow{10}{*}{\rotatebox{90}{Metadata}}  & MFR & manufacturer name from the file header\\
 & MFS & magnetic filed strength from the file header\\
 & VRX& voxel resolution in x plane\\
 & VRY & voxel resolution in y plane\\
 & VRZ & voxel resolution in z plane\\
 & ROWS & rows value of the volume \\
 & COLS & columns value of the volume\\
& TR & repetition time value of the volume\\
 & TE & echo time value of the volume\\
 & NUM& number of slice images in each volume\\
\Xhline{2\arrayrulewidth}
 & MEAN & mean of the foreground ($m_1=\mu_F = \frac{1}{MN}\sum_{i,j}F(i,j)$)\\
\multirow{25}{*}{\rotatebox{90}{Measurements}} & RNG & range of the foreground ($m_2 = {\rm{max}}(F) - {\rm{min}}(F)$) \\
 & VAR & variance of the foreground ($m_3 = \sigma^2_F$)\\
 & CV &  coefficient of variation of the foreground for shadowing and inhomogeneity artifacts~\cite{doi:10.1002/mp.13245} ($m_4 = \frac{\sigma_F}{\mu_F}$) \\
 & CPP & contrast per pixel~\cite{chang2015reference}: mean of the foreground filtered by a $3 \times 3$ 2D Laplacian kernel for shadowing artifacts  
 ($m_5 = {\rm{mean}}(\rm{conv2}(F,f_1)), \  f_1 = \frac{1}{8} \scriptsize{\begin{bmatrix}
-1 & -1 & -1\\
-1 & 8 & -1\\
-1 & -1 & -1
\end{bmatrix}}$) \\
& PSNR& peak signal to noise ratio of the foreground~\cite{http://bigwww.epfl.ch/publications/sage0303.html} ($m_6 = 10 {\rm{log}}\frac{{\rm{max}}^2(F)}{{\rm{MSE}}(F,f_2)}, \ f_2$ is a $5 \times 5 $ median filter) \\
& SNR1 & foreground standard deviation (SD) divided by background SD~\cite{Bushberg} ($m_7 = \frac{\sigma_F}{\sigma_B}$) \\
 & SNR2 & mean of the foreground patch divided by background SD~\cite{esteban2017mriqc}.  ($m_8 = \frac{\mu_{FP}}{\sigma_B}$) \\
 & SNR3 & foreground patch SD divided by the centered foreground patch SD ($m_9 = \frac{\mu_{FP}}{\sigma_{FP - \mu_{FP}}}$)\\
 &SNR4 & mean of the foreground patch divided by mean of the background patch ($m_{10} = \frac{\mu_{FP}}{\sigma_{BP}}$) \\
 & CNR & contrast to noise ratio for shadowing and noise artifacts~\cite{Bushberg}: mean of the foreground and background patches difference divided by background patch SD ($m_{11}= \frac{\mu_{FP - BP}}{\sigma_{BP}}$)\\
& CVP & coefficient of variation of the foreground patch for shading artifacts: foreground patch SD divided by foreground patch mean ($m_{12} = \frac{\sigma_{FP}}{\mu_{FP}}$) \\
& CJV &  coefficient of joint variation between the foreground and background for aliasing and inhomogeneity artifacts~\cite{hui2010fast} ($m_{13} = \frac{\sigma_{F} + \sigma_{B}}{\left|\mu_{F} - \mu_{B}\right|}$) \\
 & EFC & entropy focus criterion for motion artifacts~\cite{esteban2017mriqc}: \newline $m_{14} = \frac{NM}{\sqrt{NM}}\log{\frac{E}{\sqrt{NM}}}, \ E = -\sum_{i,j}\frac{F(i,j)}{F_{\rm{max}}}\ln\frac{F(i,j)}{F_{\rm{max}}}, \ F_{\rm{max}} = \sqrt{\sum_{i,j}F^2(i,j)}$\\
 &FBER & foreground-background energy ratio for ringing artifacts~\cite{shehzad2015preprocessed}: ($m_{15} = \frac{{\rm{median}}(\left|F\right|^2)}{{\rm{median}}(\left|B\right|^2)}$) \\
\Xhline{2\arrayrulewidth}
\end{tabular}
\begin{center}
\begin{tablenotes}\scriptsize
\item[*]  All the computed measurements (numbered $m_1-m_{15}$) are average values over the entire volume, which calculated for every single slice separately. Variables $M, N, F, B, FP, BP$ stand for slice width size, slice height size, foreground image intensity values, background image intensity values, foreground patch, and background patch respectively. Operators $\mu, \sigma, \sigma^2, {\rm{median}}$ stand for mean, standard deviation (SD), and variance measures respectively. Foreground and background patches are random $5\times 5$ square patches of the foreground and background images respectively. Measurements SNR1-SNR4 are all signal to noise ratio with different definitions. 
\end{tablenotes}
\end{center}
\end{threeparttable}
}
\end{center}
\end{table}


\subsection{Front-end Visualization}
\sethlcolor{reviewer2}\hl{The interactive user-interface of MRQy was built as a locally hosted HTML5/Javascript file (compatible with popular web browsers such as Google Chrome, Firefox, or Chromium); specifically designed for real-time analytics, data filtering, and interactive visualization by the end-user.} The goal was to allow end-users to easily investigate trends in site or scanner variations within an MRI cohort, as well as identify those scans that require additional processing due to the presence of artifacts. The MRQy interface splits into 3 major sections, all of which are inter-connected. Any individual section can also be disabled or re-enabled by the end-user to provide a fully customizable interface.

\subsubsection{Tables}
Extracted metadata and measures appear in separate tables within the interface. Each table has sortable columns to easily view outliers in numeric values. Any incomplete metadata values are displayed as ``NA'' (as well as being ignored in subsequent visualizations). Information can easily be copied out of the tables, which are also fully configurable including allowing for removal of specific subjects or specific columns. 

\subsubsection{Charts}
\begin{packed_item}
\item {\it\underline{{{\sethlcolor{reviewer1}\hl{Parallel Coordinate (PC):}}}}}
This is a multivariate visualization tool which has been shown to be effective for understanding trends within multi-variate datasets~\cite {edsall2003parallel}. For each MRI volume, a polyline is plotted (i.e. an unbroken line segment) which connected vertices of patient measurements on the parallel axes; i.e the vertex position on each axis corresponds to the value of the point for that specific metadata field or measure. The PC chart offered a visual approach to check variations in each measure as they related to the rest of the cohort as well as aided in the identification of outliers. 

\item {\it\underline{{Bar:}}}
A single bar is plotted per MRI volume for a selected variable (either metadata or measures). This provides an alternative approach to evaluating individual variables, where outliers would be markedly taller or shorter than the remainder of the cohort. 

\end{packed_item}


\subsubsection{Scatter Plots}
To examine how the MRI volumes in a cohort relate to one another as well as to examine site- or scanner-specific trends, 2 different ``embeddings'' were computed based on the t-SNE~\cite{maaten2008visualizing} and UMAP~\cite{ mcinnes2018umap} algorithms within the Python backend.
Both these algorithms take as an input all 23 measures for each patient and output a 2-dimensional embedding space (visualized as a scatter plot) where similarities between patients are preserved. While t-SNE yields a relatively robust representation of overall cohort structure, UMAP additionally provides a topological data structure that has been shown to be more accurate than t-SNE in some settings~\cite{ mcinnes2018umap}.
t-SNE was implemented using the  \verb|sklearn.manifold| Python package~\cite{pypi} with default parameters (e.g. \verb|n_components = 2, perplexity = 30, random_state = 0|), while UMAP utilized the \verb|umap-learn| Python package~\cite{pypi} in the backend with default parameters (\verb|n_components = 1, n_neighbors= 15, min_dist= 0.1, metric = `euclidean'|). 


\subsection{Experimental Design}

\subsubsection{TCGA-GBM cohort}
TCGA-GBM is the largest available dataset of brain MRI data from the Cancer Imaging Archive (TCIA)~\cite{tcia}, and comprises scans from n=259 subjects. This study was limited to subjects for whom a {\sethlcolor{reviewer1}\hl{post-contrast T1-weighted MR image}} in the axial plane was available, resulting in a cohort of n=133 T1-POST MRI scans accrued from 7 different sites (also visualized in Figure \ref{fig_fore_back}). As these MRI scans were acquired under different environmental conditions and using different scanner equipment and imaging protocols, this cohort includes typical data variations and image artifacts that may be observed in a TCIA dataset~\cite{bakas2017advancing}. All 133 T1-POST studies were downloaded as DICOM files from TCIA.  \par 
\hlfr{In addition to the original cohort (as downloaded from TCIA), a processed version of the TCGA-GBM cohort was obtained from a publicly accessible release by Bakas et al{\cite{bakas2017advancing}}. Briefly, processed MR volumes had been made available as NIFTI files after undergoing the following steps: (1) re-orientation to the left-posterior-superior coordinate system, (2) co-registration to the T1w anatomical template of SRI24 Multi-Channel Normal Adult Human Brain Atlas via affine registration{\cite{Beigclincanres.2556.2019}}, (3) resampling to $1\ mm^3$ voxel resolution{\cite{rohlfing2010sri24}}, (4) skull-stripping{\cite{davatzikos2018cancer}}, (5) de-noising using a low-level image processing smoothing filter{\cite{smith1997susan}}, and intensity standardization to an image distribution template{\cite{nyul2000stdn}}}. 
\par 
\subsubsection{In-house rectal cancer cohort}
This retrospectively curated dataset comprised 104 patients accrued from three different institutions including the Cleveland Clinic (CC, n=60), University Hospitals (UH, n=35), and the Cleveland Veterans Affairs (VA, n = 9) (further details available in [\citenum{antunes2020}]). For all patients, a T2-weighted turbo spin-echo MRI sequence had been acquired from patients diagnosed with rectal adenocarcinoma. While all scans were anonymized DICOM files, they had been acquired with different sequence parameters and using different scanner equipment; thus making the cohort an exemplar of a retrospectively accrued multi-site cohort. \par 
\hlfr{Both the original scans as well as processed scans in this cohort were utilized, where in the latter case all MRI datasets had undergone the following sequence of operations: (1) linear resampling to a consistent voxel resolution of $0.781 \times 0.781 \times 4.0 \  mm$, (2) N4ITK bias field correction{~\cite{tustison2010n4itk}}, and (3) image intensity normalization with respect to a muscle region{~\cite{Sunoqrot2020, Scalco2020}}. The processed images were available as MHA files.}

\subsubsection{\colorbox{reviewer1}{Evaluation}}
\hlfr{The MRQy front-end was used to assess each of the two cohorts for batch effects as well as image quality artifacts. In order to determine how congruent and consistent the cohorts were as a result of processing, the output of MRQy was compared between the original unprocessed and the processed scans (i.e. after applying artifact correction and normalization operations). All analysis was conducted using Python 3.7.4 for backend processing and Google Chrome 80.0.3987.149 for interacting with the front-end on a PC with Intel(R) Core(TM) i7 CPU 930 (3.60 GHz), 32 GB RAM, and a 64 bit Windows 10 operating system.}
\sethlcolor{reviewer1}\hl{
The performance of MRQy measures in identifying batch effects in each cohort without any \textit{a priori} knowledge was evaluated via consensus clustering{\cite{monti2003consensus}} of all 23 measures using the} \verb|ConsensusClusterPlus| \sethlcolor{reviewer1}\hl{package in R{\cite{wilkerson2010consensusclusterplus}}. Hierarchical consensus clustering (with $k=7$ for TCGA-GBM and $k=3$ for the rectal cohort) was performed using Pearson distance and 1000 iterations, including 80\% random dataset resampling between iterations. The result was visualized as a consensus cluster heatmap where the shading indicated the frequency with which a pair of patients was clustered together. Cluster overlap accuracy was then computed by first identifying which cluster corresponded to which site (based on precision/recall values) and then calculating what fraction of the datasets from each site had been correctly clustered together in consensus clustering.
}

\section{Results}

\subsection{Assessment of TCGA-GBM datasets for batch effects and imaging artifacts}
\begin{figure}[t!]
\begin{center}
\includegraphics[width=1\linewidth]{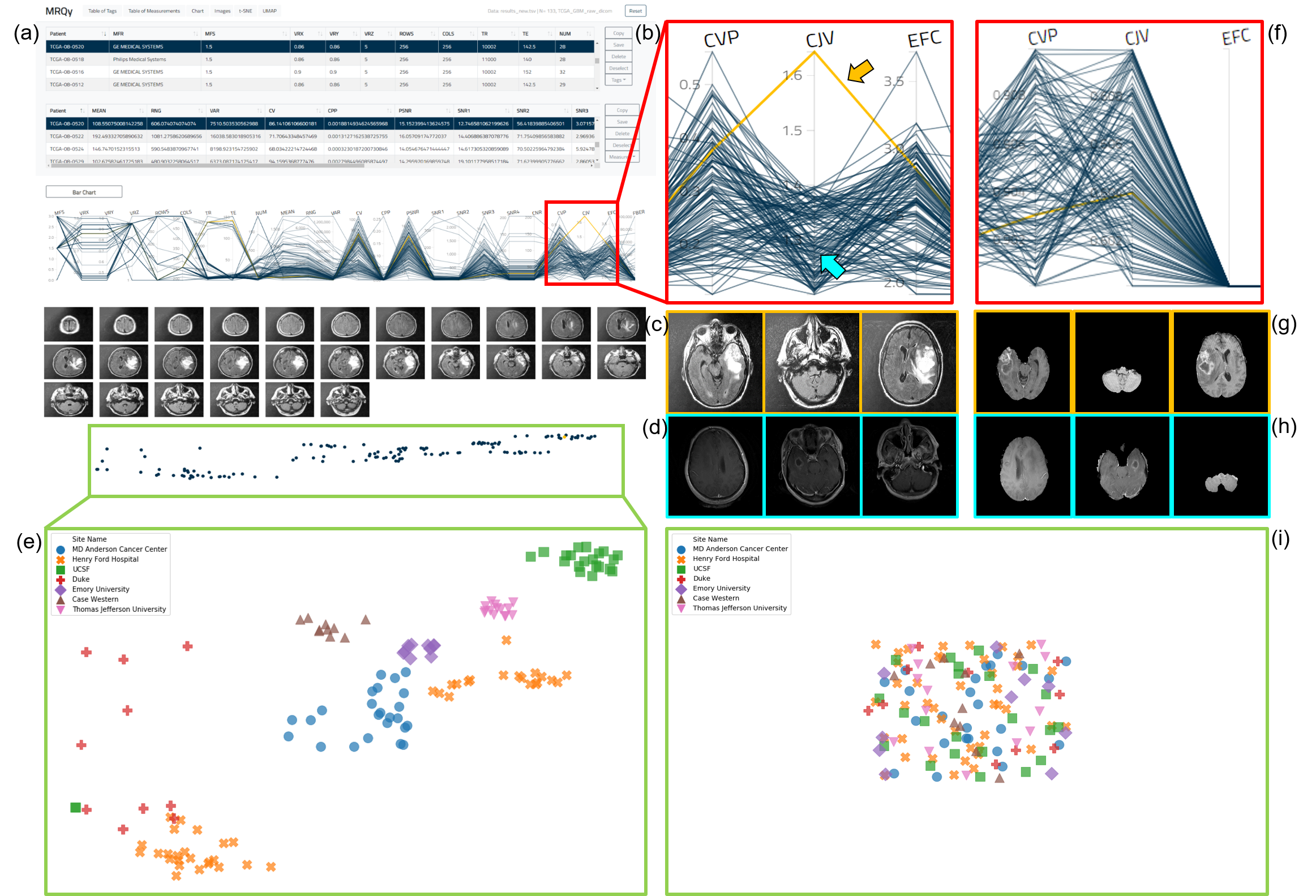}
\captionv{15}{}
{(a) MRQy front-end interface for interrogating TCGA-GBM cohort before processing. (b) Outlier dataset identified on PC chart for the CJV measure found to exhibit shading artifact on (c) representative images, especially when compared to (d) a different dataset. (e) Scatter plot revealing presence of site-specific batch effects in this cohort before processing (colored symbols corresponding to different sites appear in site-specific clusters). (f) corresponds to (b) after data processing. (g) Processed images corresponding to (c). (h) Processed images corresponding to (d). (i) Scatter plot of the processed data fall with in a single merged cluster.
\label{gbm_result}}  
\end{center}
\end{figure}

\sethlcolor{reviewer1}\hl{Analysis via MRQy took  93.5 minutes to process all 133 datasets in the original TCGA-GBM cohort ($\approx42$s/dataset) and 101.6 minutes ($\approx46$s/dataset) for the processed cohort.}
Quality control of the TCGA-GBM cohort via the MRQy front-end interface (Figure \ref{gbm_result}(a)) reveals the following: 
\begin{itemize}
    \item \underline{\textit{Imaging artifact detection}:} The PC chart of the CJV measure (Figure \ref{gbm_result}(b), unprocessed cohort mean CJV=$1.30\pm0.06$) shows the presence of a distinct outlier (highlighted in orange + orange arrow, CJV=1.64). Further visualizing representative images (Figure \ref{gbm_result}(c)) from this dataset depicts a distinct shading artifact compared to a different dataset from the cohort ((Figure \ref{gbm_result}(c), cyan arrow in PC chart, CJV=1.2). This is the specific image artifact that the CJV measure is designed to  quantify~\cite{hui2010fast}, which indicates this dataset requires bias correction or intensity normalization prior to computational analysis.
    \item \underline{\textit{Correction of image artifacts after processing}:} \hlfr{Examining the PC chart of the CJV measure of processed scans (Figure {\ref{gbm_result}}(f), cohort mean CJV=$0.605\pm0.05$) reveals no distinct outliers (the highlighted dataset from Figure {\ref{gbm_result}}(b) has CJV=0.604 after processing). Figure {\ref{gbm_result}}(g) and (h) depict processed images corresponding to those in Figure {\ref{gbm_result}}(c) and (d) respectively. This suggests that the processing steps have accounted for the intensity artifacts previously identified, as reflected in the CJV measure.}
    \item \underline{\textit{Batch effect detection}:} Significant site-specific variations are seen to be present in the embedding scatter plot overlaid with different colored symbols for each site ((Figure \ref{gbm_result}(e)). Each of the 7 sites appear as a distinct cluster, suggesting batch effects that need to be corrected for prior to model development. 
    \item \underline{\textit{Correction of batch effects after processing}:} \hlfr{The embedding scatter plot reveals no distinctive clusters corresponding to any site, where the colored symbols in Figure {\ref{gbm_result}}(i) fall within a single merged cluster; suggesting that the processing steps may have successfully accounted for site-specific variations.}
\end{itemize}

\captionsetup[sub]{font=small,labelfont=small}

\subsection{Evaluation of multi-site rectal cancer cohort via MRQy}
\captionsetup[sub]{font=small,labelfont=small}
\begin{figure}[t!]
\begin{center}
\includegraphics[width=1\linewidth]{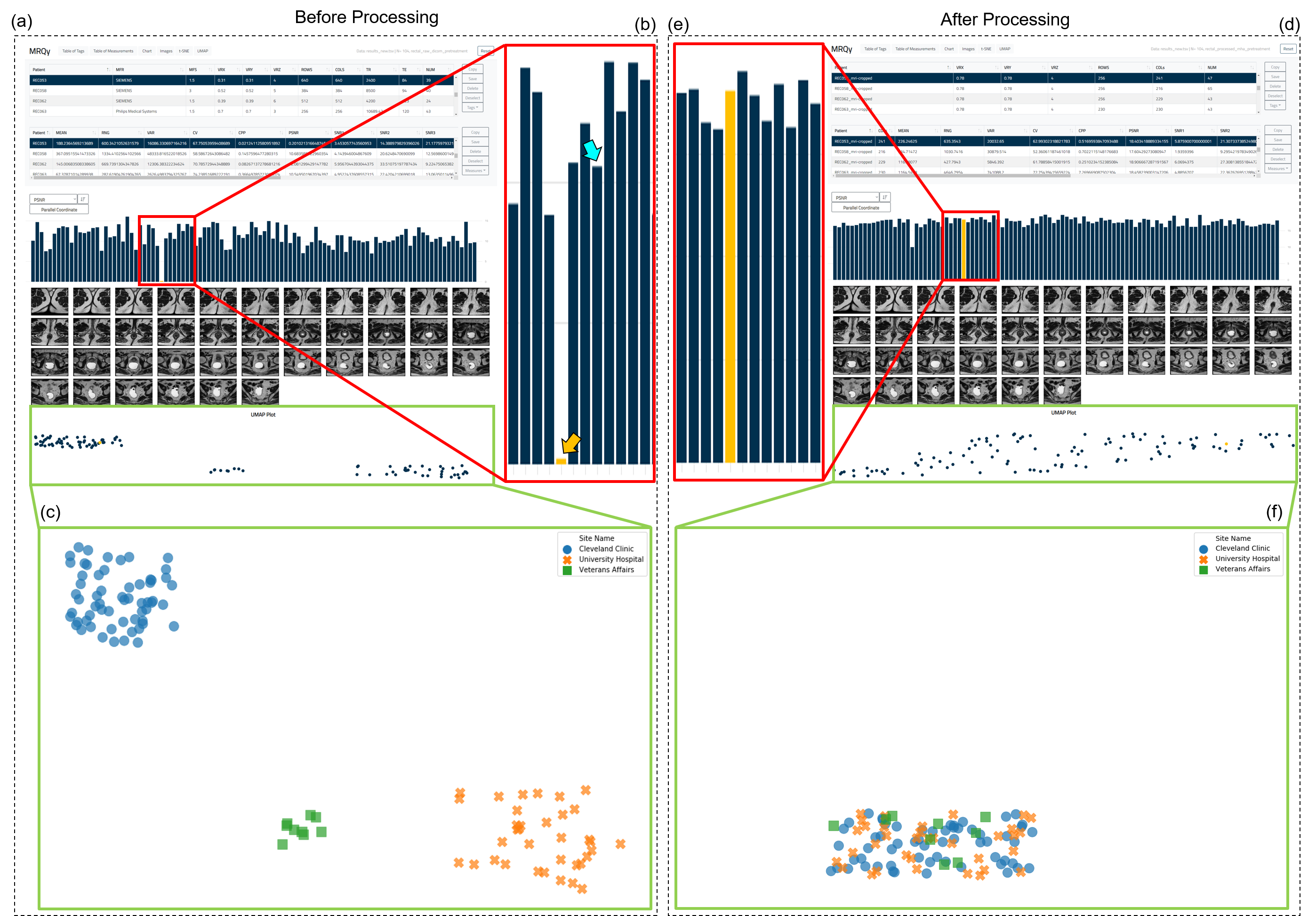}
\captionv{15}{}
{(a) MRQy front-end interface for interrogating rectal cohort before processing. (b) Outlier dataset identified on the bar chart for the PSNR measure found to exhibit subtle noise artifacts. (c) Scatter plot reveals presence of site-specific clusters in this cohort before processing (colored symbols corresponding to different sites appear in site-specific clusters). (d) MRQy front-end interface showing results using processed data. (e) Improved PSNR value of dataset highlighted in (b) after processing, also more consistent with remainder of cohort. (f) Scatter plot a single merged cluster for the cohort, suggesting batch effects and site-specific variations may have been accounted for.  
\label{rectal_result}}  
\end{center}
\end{figure}

\hlfr{Analysis via MRQy took 51.8 minutes to process all 104 original datasets in rectal cohort ($\approx30$s/dataset) and 60.28 minutes ($\approx35$s/dataset) for the processed cohort. Using the MRQy front-end (Figure {\ref{rectal_result}}(a) for raw dataset and Figure {\ref{rectal_result}}(d) for processed data) for quality control revealed the following:}
\begin{itemize}
    \item \underline{\textit{Imaging artifact detection before and after processing}:} \hlfr{The bar chart of PSNR values (Figure {\ref{rectal_result}}(b)), unprocessed cohort mean PSNR=$11.17\pm2.38$) helps quickly identify a specific dataset with very poor PSNR (indicated via orange arrow, PSNR=0.2). PSNR thus appears to accurately characterize noise artifacts in MR images as designed{\cite{esteban2017mriqc}}, suggesting this dataset requires denoising prior to analysis. In the processed cohort, the PSNR value for this dataset (Figure {\ref{rectal_result}}(e), orange arrow, PSNR=18.4) is markedly higher as well as appears more consistent  with the remainder of the cohort (processed cohort mean PSNR=$17.94\pm1.42$).}
    \item  \underline{\textit{Batch effect detection before and after processing}:}  Marked batch effects are seen to be present in the scatter plot in Figure \ref{rectal_result}(e) as each of the colored symbols (corresponding to different sites in the unprocessed cohort) appear in distinct clusters. \hlfr{By contrast, the scatter plot of the processed cohort in Figure {\ref{rectal_result}(f)} reveals a single merged cluster suggesting that site-specific differences have been accounted for.}

\end{itemize}

\subsection{\colorbox{reviewer1}{Quantitative evaluation of batch effects before and after processing}}
\begin{figure}[!ht]
\vspace{0.41cm}
\begin{center}
\begin{subfigure}[b]{0.49\textwidth}
\resizebox {1\textwidth}{!}{
\includegraphics[width=5.5cm,height=4.5cm]{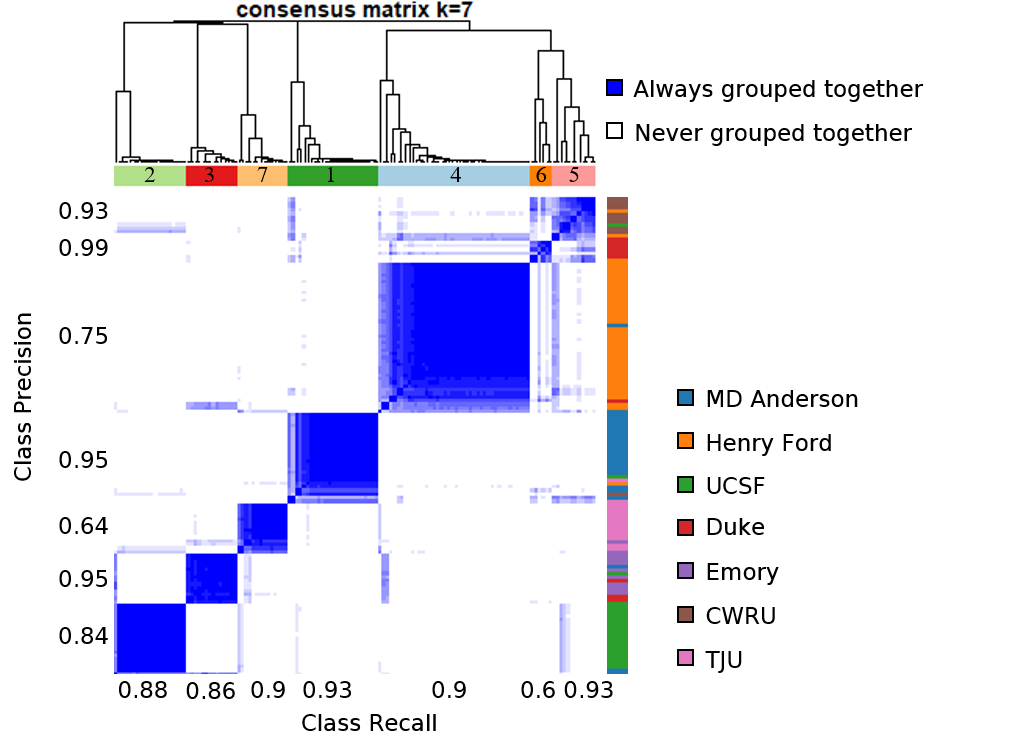}
}
\caption{}
\label{cc1}
\end{subfigure}
\hspace{-0.7cm}
\begin{subfigure}[b]{0.49\textwidth}
\resizebox {1\textwidth}{!}{
\includegraphics[width=5.5cm,height=4.5cm]{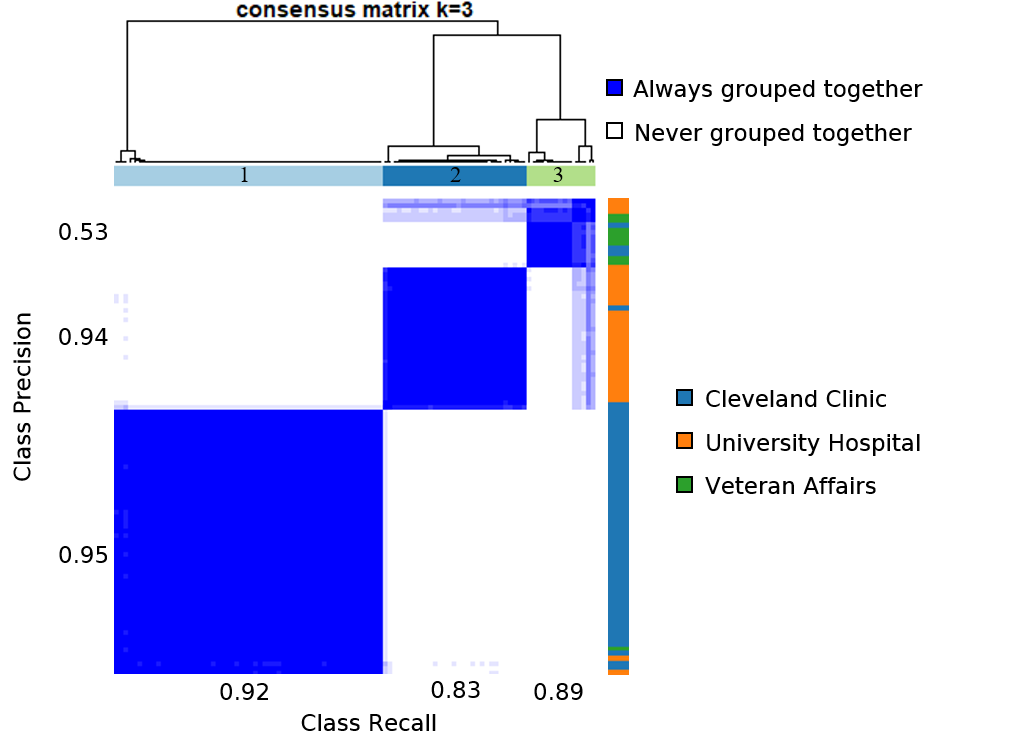}
}
\caption{}
\label{cc2}
\end{subfigure}\\
\begin{subfigure}[b]{0.49\textwidth}
\resizebox {1\textwidth}{!}{
\includegraphics[width=5.5cm,height=4cm]{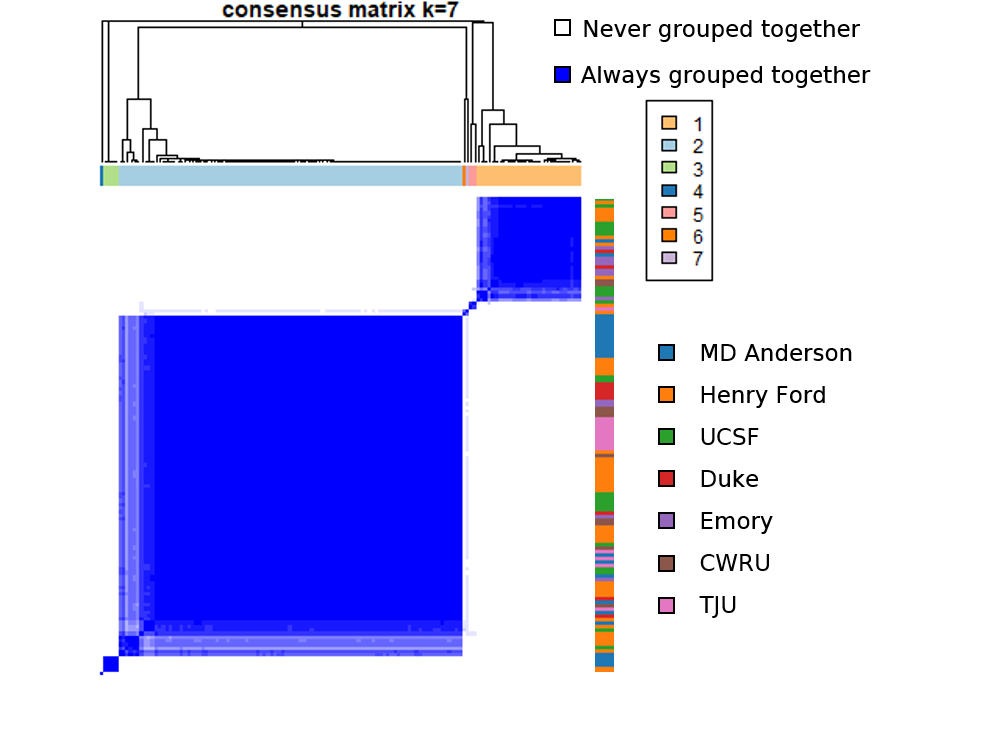}
}
\caption{}
\label{process3}
\end{subfigure}
\hfill
\begin{subfigure}[b]{0.49\textwidth}
\resizebox {1\textwidth}{!}{
\includegraphics[width=5.5cm,height=4cm]{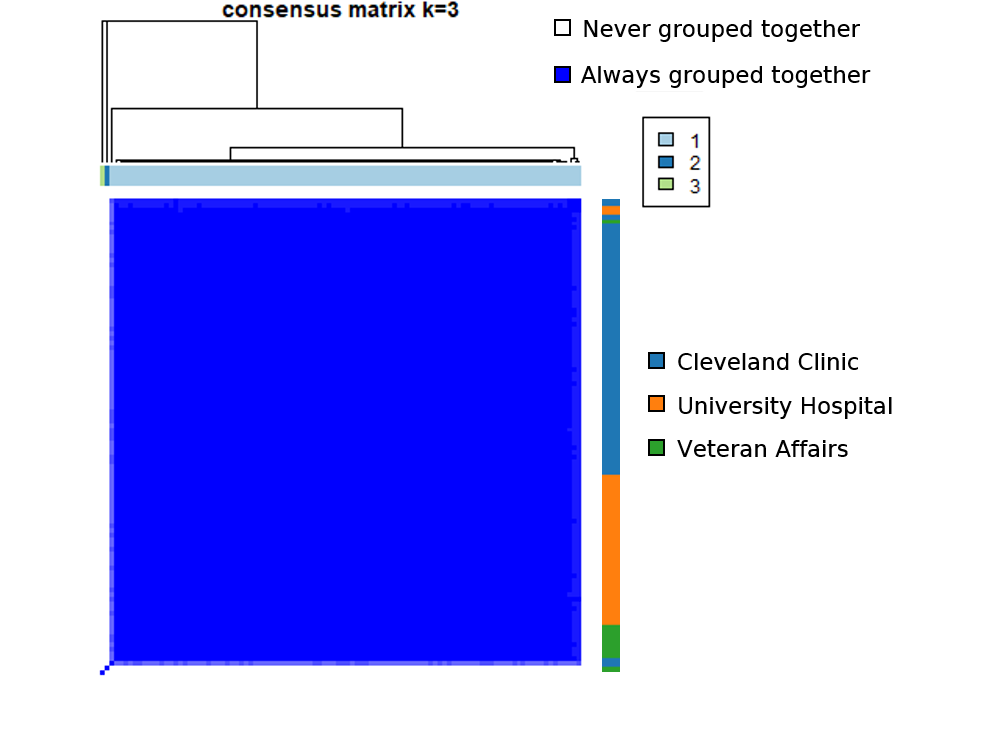}
}
\caption{}
\label{process4}
\end{subfigure}
\captionv{15}{}
{\sethlcolor{reviewer1}\hl{Consensus clustegram of all 23 MRQy measures computed for (a) unprocessed TCGA-GBM cohort, (b) unprocessed rectal cohort, (c) processed TCGA-GBM cohort, and (d) processed rectal cohort. In each plot, the colorbar along the right corresponds indicates which site each dataset belongs to (legend alongside) while the colorbar at the top is the cluster label obtained via consensus clustering. For unprocessed data, precision and recall values for each cluster in identifying a specific site are also noted along the left and bottom of the clustergram, respectively.}   
\label{conf_mat}}  
\end{center}
\end{figure}

\hlfr{Figure {\ref{conf_mat}}(a) and (b) visualizes the clustergrams for each cohort (based on unprocessed data) obtained via consensus clustering of the 23 measures. For the TCGA-GBM cohort, the cluster overlap accuracy of MRQy measures with respect to each of the 7 sites is 87.5\%, 86.4\%, 90\%, 93\%, 90\%, 60\%, and 92.9\% respectively. Similarly, MRQy measures are found to cluster each of the 3 sites in the rectal cohort with an overlap accuracy of 91\%, 82.8\%, and 88.9\%, respectively.
The strong co-clustering of datasets from the same site as as well as relatively high clustering accuracy of datasets within site-specific groupings are suggestive of batch effects in both cohorts.}

\sethlcolor{reviewer1}\hl{Corresponding clustergrams for the processed datasets are shown in Figure {\ref{conf_mat}(c)} (for TCGA-GBM cohort) and Figure {\ref{conf_mat}(d)} (for rectal cohort). After processing, the cluster overlap accuracy of MRQy measures in the TCGA-GBM cohort for each of the 7 sites reduced to 16.67\%, 2.33\%, 4.55\%, 70\%, 10\%, 20\%, and 7.14\% respectively. Similarly, the cluster overlap accuracy for each of the 3 sites in the rectal cancer cohort is 1.67\%, 0\%, and 88.89\%, respectively. This suggests that none of the clusters identified via MRQy measures are associated with any specific site after processing, in either cohort.}  

\captionsetup[sub]{font=small,labelfont=small}
\section{Discussion}
We have presented a technical overview and experimental evaluation of a novel open-source quality control tool for MR imaging data; called MRQy.
Using cohorts from 2 different body regions (brain MRIs from TCIA, in-house rectal MRIs), we demonstrated that MRQy can quantify and evaluate (a) site- or scanner-specific variations within an MRI cohort, (b) imaging artifacts that compromise relative image quality of MRI datasets, and (c) how well specific processing pipelines have accounted for artifacts that are present in an given data cohort.

\hlfr{As each of the metadata and quality measures extracted within MRQy can be used to reveal a variety of different artifacts that may be present in MR images, we opted to examine two representative exemplar artifacts in this work.
Our results demonstrate that we were able to accurately identify datasets with specific intensity and noise artifacts in this fashion, as they appear as distinct outliers in our specialized interface.
More importantly, by iteratively using MRQy to QC the cohorts, we were able to determine how well these artifacts were corrected in the processed MRI datasets (i.e. after artifact correction and image normalization operations were performed).
We additionally conducted an extensive qualitative and quantitative evaluation for the presence of batch effects in both cohorts, both before and after processing.
While the presence of batch effects and their impact on IQMs have been evaluated in previous studies{~\cite{esteban2017mriqc}} from a supervised learning perspective, we used a consensus clustering approach to quantify how well specific sites could be identified in the cohort in a completely unsupervised fashion. This provided us a purely data-driven indication of how well the processing pipelines have corrected the site- and scanner-specific variations present in both cohorts considered in this work.
While MRQy was used to evaluate MRI datasets processed using specific pipelines, our tool could be used to evaluate the impact of individual steps within the pipeline as thus determine the optimal sequence of processing operations that should be applied to a given MRI cohort.}

Quality control of MR imaging has long been a topic of active research~\cite{Ieremeiev2018, woodard2006no,Mortamet2009,Pizarro2016,ding2019supervised}, but a majority of the resulting tools have been specifically developed for brain MRIs~\cite{esteban2017mriqc,KLAPWIJK2019116,vogelbacher2019lab} in order to identify acceptable scans via supervised learning (based on subjective expert quality ratings). By contrast, MRQy represents a unique solution for quality control of MRI datasets that has been designed to work with scans of any body region while running in an efficient and unsupervised fashion. MRQy has been developed by expanding on an initial framework developed as part of a digital pathology quality control tool, HistoQC~\cite{janowczyk2019histoqc}, but has been evolved significantly to specialize it for MR imaging data by incorporating imaging-specific processing steps and quality measures. Unlike existing approaches for MRI quality control, MRQy also includes a specialized interactive front-end which can help quickly identify what relative imaging artifacts and batch effects are present in a cohort. 
While the current work has been limited to demonstrating detection of 2 specific artifacts (shading, noise) via MRQy, we believe this approach will generalize to other artifacts due to the extensive list of extracted quality measures which have been previously validated for this purpose~\cite{esteban2017mriqc}.

The recent focus on the reproducibility of radiomics and deep learning across sites~\cite{chirra2019multisite} implies a need to interrogate cohort differences which can cause models trained on one site to fail to generalize on another. 
To enable this, we have released results of our QC analysis of MRI studies from the TCGA-GBM as well as 2 other TCIA collections (\url{https://doi.org/10.7937/K9/TCIA.2020.JHZ2-T694}) which could help researchers in addressing these questions for large-scale multi-site or multi-scanner cohorts~\cite{um2019impact,moradmand2019impact}.
We envision using MRQy as a pre-analytic step in any computational imaging pipeline to determine (a) what site- or scanner-specific variations need to be accounted for to accurately characterize ``true'' biological differences, as well as (b) which corrections should be applied to curate datasets of acceptable relative quality (i.e. minimal or no outliers); prior to model development or validation.
\par 
Further development of MRQy has been enabled by its modular design, exemplified by the HTML front-end which is already highly customizable by the end-user. One of our planned functionalities is to enable real-time generation of the embedding scatter plots directly within the MRQy HTML front-end rather than via the backend. Similarly, the Python backend can be easily modified via plugins to use a different foreground detection algorithm, compute additional quality measures, or even to integrate quality prediction algorithms~\cite{ding2019supervised,osadebey2018standardized} in the future.
One of the current limitations of MRQy is that it can only be used for quality control of structural MRI data, but we are working on developing new measures that could allow MRQy to be used to interrogate non-structural MRIs (e.g. diffusion or {\sethlcolor{reviewer1}\hl{dynamic contrast-enhanced MRI}}).
Additionally, by modifying the quality measures appropriately, this tool could be eventually be adapted for quality control of any radiographic imaging collection (CT, ultrasound, PET, etc.) that is housed in public repositories such as TCIA. 
\section{Conclusions}
We have presented a technical overview and experimental evaluation of MRQy, a new quality control tool for MR imaging data. MRQy works by computing actionable quality measurements as well as image metadata from any structural MR sequence of any body region via a Python backend, which can be interrogated via  a specialized HTML5 front end. MRQy can be executed completely offline and the entire tool requires only a few commands to run in an unsupervised fashion, as well as being designed to be easily extensible and modular. Our initial results successfully demonstrated how MRQy could be used for (a) quantifying site- and scanner-specific batch effects within large multi-site cohorts of MR imaging data (such as TCIA), (b) identifying relative imaging artifacts within MRI datasets which require correction prior to model development, as well as (c) evaluating how well specific processing pipelines have corrected for these issues in a given data cohort. MRQy has been made publicly accessible as an open-source project through GitHub (\url{https://github.com/ccipd/MRQy}), and can be downloaded as well as contributed to freely by any end-user. 

\section*{Acknowledgements}
Research reported in this publication was supported by the National Cancer Institute of the National Institutes of Health under award numbers 1U24CA199374-01, R01CA202752-01A1, R01CA208236-01A1, R01CA216579-01A1, R01CA220581-01A1, 1U01CA239055-01, 1F31CA216935-01A1, 1U01CA248226-01, National Institute for Biomedical Imaging and Bioengineering 1R43EB028736-01, National Center for Research Resources under award number 1 C06 RR12463-01, VA Merit Review Award IBX004121A from the United States Department of Veterans Affairs Biomedical Laboratory Research and Development Service, the DoD Breast Cancer Research Program Breakthrough Level 1 Award (W81XWH-19-1-0668), the DOD Prostate Cancer Idea Development Award (W81XWH-15-1-0558), the DOD Lung Cancer Investigator-Initiated Translational Research Award (W81XWH-18-1-0440), the DOD Peer Reviewed Cancer Research Program (W81XWH-16-1-0329, W81XWH-18-1-0404), the Kidney Precision Medicine Project (KPMP) Glue Grant, the Dana Foundation David Mahoney Neuroimaging Program, the V Foundation Translational Research Award, the Ohio Third Frontier Technology Validation Fund, the Wallace H. Coulter Foundation Program in the Department of Biomedical Engineering and The Clinical and Translational Science Award Program (CTSA) at Case Western Reserve University.

The content is solely the responsibility of the authors and does not necessarily represent the official views of the National Institutes of Health, the U.S. Department of Veterans Affairs, the Department of Defense, or the United States Government.

\section*{Supplementary Materials}
\addcontentsline{toc}{section}{\numberline{}Supplemental Materials}

\subsection*{\colorbox{reviewer1}{Definitions}}
\hlfr{\textit{Batch  effects}  refer to systematic occurrence of a technical artifact within a subset of a cohort{\cite{leek2010tackling}}. For example, the slice spacing may be different between two institutions, creating a batch effect if those studies were evaluated in unison. Consider the following example. Site A produces MR volumes with a larger slice spacing as well as a majority of the cases belonging to the positive target class, while Site B provides MR volumes with much smaller slice spacing and a majority of cases from the negative class. When combining data from both sites, any developed model must be optimized to identify biologically relevant features specific to the target class, as opposed to determining that slice spacing may be relevant for distinguishing between the 2 classes (as the latter is really an acquisition difference between the 2 sites).

\textit{Image artifacts} refer to the image quality issues such as the presence of noise, motion, shading, lack of detail, ringing, aliasing or other issues; that are present within individual MR datasets to varying degrees{\cite{VanEssen2012}}. These typically occur as a result of the acquisition process, scanner calibration issues, or technician error.}

\subsection*{Format and Usage}
\label{FAU}
MRQy has been made publicly accessible as an open-source project through GitHub (\url{https://github.com/ccipd/MRQy}), and can be downloaded as well as contributed to freely by any end-user. Specific package dependencies for MRQy have included in the GitHub documentation. After the installation of all the prerequisite Python packages (specified in the installation instructions), MRQy can be run on a directory containing files for a given cohort via the command: \verb|python QC.py output_folder_name "input directory address”|. No additional configuration files need to be specified. This results in the following steps being executed:
\begin{packed_item}
    \item Thumbnail images are generated for all 2D sections in each MRI dataset and saved as {\it{.png}} files within the \verb|UserInterface/Data| folder.
    \item Each dataset is processed to detect the foreground and background region.
    \item \textit{Metadata} are extracted from file headers for each dataset. \textit{Measurements} are computed based on the detected foreground region for each dataset.
    \item Both metadata and measurements are saved for each dataset within a tab-separated file ({\it{results.tsv}}) that is stored within the \verb|UserInterface/Data| folder.
    \item For a given cohort, a single UMAP and a single t-SNE embedding are computed for all the datasets based on the 23 measures (after whitening). The embedding co-ordinates are also saved into the {\it{results.tsv}} file.
\end{packed_item}
Further interrogation of cohort variations and artifact trends may be done reading  {\it{results.tsv}} into any common data analytic tool (e.g. MATLAB or R). A specialized front-end HTML interface ({\it{index.html}}) is available within the \verb|UserInterface| folder designed for real-time manipulation and visualization. Quality control can be performed via multiple pathways:
\begin{packed_item}
\item 
Using sorting arrows available on each table column to re-order measures and examine numeric trends. Users can further annotate rows or remove non-informative patients.
\item 
As the different interface components are synchronized, if a patient row is highlighted in either Table, a corresponding highlight appears on a line within the PC chart, on a bar in the bar chart, as well as shading the patient-specific bubble in the embedding plots. Thumbnail images for this patient volume are shown in the interface. 
\item 
Using the PC and bar charts to directly compare a specific measure across all the subject scans. This can help quickly determine which of the metadata or measures are consistent across the entire cohort as well as identify outliers. The PC chart can also be used to evaluate positive or negative relationships between different measures~\cite{636793} and thus determine the trade-off in processing for specific artifacts.
\item 
Using embedding plots (t-SNE and UMAP) to track specific site- or scanner-specific trends within the cohort. By visualizing the 2D space into which the entire cohort has been mapped, any clusters that can be identified typically correspond to site- and scanner-specific variations. The overall distribution of points in space also provide an indication of the variability within the entire cohort. 
\end{packed_item}

\subsection*{\colorbox{reviewer1}{User-Specified Settings}}
\hlfr{By default, MRQy extracts a basic set of 10 metadata tags from {\textit{.dcm}} files, summarized in Table {\ref{table}}. Additional tag fields or private metadata can also be extracted by specifying them via a {\textit{.txt}} file and using the syntax:} 
\newline {\color{black}
\verb|python QC.py output_folder_name "input directory” -t "tags .txt address”|.}
\par 
\hlfr{The user can also specify a configuration for the foreground detection algorithm using the following command: } 
\newline {\color{black}
\verb|python QC.py output_folder_name "input directory” -c "False”|.}
\par
\hlfr{Note that the default value for the} \verb|-c| \hlfr{flag is {\textit{False}}, where MRQy measures are computed across all foreground objects together. When the} \verb|-c| \hlfr{flag is {\textit{True}}, MRQy will identify each foreground object separately and compute a measurement per individual object.}

\section*{References}
\addcontentsline{toc}{section}{\numberline{}References}

\begin{thebibliography}{10}

\bibitem{clark2013cancer}
K.~Clark et~al.,
\newblock The Cancer Imaging Archive (TCIA): maintaining and operating a public
  information repository,
\newblock Journal of digital imaging {\bf 26}, 1045--1057 (2013).

\bibitem{prior2017public}
F.~Prior, K.~Smith, A.~Sharma, J.~Kirby, L.~Tarbox, K.~Clark, W.~Bennett,
  T.~Nolan, and J.~Freymann,
\newblock The public cancer radiology imaging collections of The Cancer Imaging
  Archive,
\newblock Scientific data {\bf 4}, 170124 (2017).

\bibitem{kalpathy2014quantitative}
J.~Kalpathy-Cramer, J.~B. Freymann, J.~S. Kirby, P.~E. Kinahan, and F.~W.
  Prior,
\newblock Quantitative imaging network: data sharing and competitive
  AlgorithmValidation leveraging the cancer imaging archive,
\newblock Translational oncology {\bf 7}, 147--152 (2014).

\bibitem{zanfardino2019tcga}
M.~Zanfardino, K.~Pane, P.~Mirabelli, M.~Salvatore, and M.~Franzese,
\newblock TCGA-TCIA Impact on Radiogenomics Cancer Research: A Systematic
  Review,
\newblock International Journal of Molecular Sciences {\bf 20}, 6033 (2019).

\bibitem{prior2020open}
F.~Prior, J.~Almeida, P.~Kathiravelu, T.~Kurc, K.~Smith, T.~Fitzgerald, and
  J.~Saltz,
\newblock Open access image repositories: high-quality data to enable machine
  learning research,
\newblock Clinical radiology {\bf 75}, 7--12 (2020).

\bibitem{basu2019call}
A.~Basu, D.~Warzel, A.~Eftekhari, J.~S. Kirby, J.~Freymann, J.~Knable,
  A.~Sharma, and P.~Jacobs,
\newblock Call for Data Standardization: Lessons Learned and Recommendations in
  an Imaging Study,
\newblock JCO Clinical Cancer Informatics {\bf 3}, 1--11 (2019).

\bibitem{schlett2016quantitative}
C.~L. Schlett et~al.,
\newblock Quantitative, organ-specific interscanner and intrascanner
  variability for 3 T whole-body magnetic resonance imaging in a multicenter,
  multivendor study,
\newblock Investigative radiology {\bf 51}, 255--265 (2016).

\bibitem{glocker2019machine}
B.~Glocker, R.~Robinson, D.~C. Castro, Q.~Dou, and E.~Konukoglu,
\newblock Machine learning with multi-site imaging data: An empirical study on
  the impact of scanner effects,
\newblock arXiv preprint arXiv:1910.04597  (2019).

\bibitem{wachinger2019quantifying}
C.~Wachinger, B.~G. Becker, A.~Rieckmann, and S.~P{\"o}lsterl,
\newblock Quantifying confounding bias in neuroimaging datasets with causal
  inference,
\newblock in {\em International Conference on Medical Image Computing and
  Computer-Assisted Intervention}, pages 484--492, Springer, 2019.

\bibitem{leek2010tackling}
J.~T. Leek, R.~B. Scharpf, H.~C. Bravo, D.~Simcha, B.~Langmead, W.~E. Johnson,
  D.~Geman, K.~Baggerly, and R.~A. Irizarry,
\newblock Tackling the widespread and critical impact of batch effects in
  high-throughput data,
\newblock Nature Reviews Genetics {\bf 11}, 733--739 (2010).

\bibitem{tcia}
L.~Scarpace, T.~Mikkelsen, S.~Cha, S.~Rao, S.~Tekchandani, D.~Gutman, J.~Saltz,
  B.~Erickson, N.~Pedano, A.~Flanders, J.~Barnholtz-Sloan, Q.~Ostrom,
  D.~Barboriak, and L.~Pierce,
\newblock Radiology Data from The Cancer Genome Atlas Glioblastoma Multiforme
  [TCGA-GBM] collection,
\newblock The Cancer Imaging Archive ,
\newblock http://doi.org/10.7937/K9/TCIA.2016.RNYFUYE9.

\bibitem{zwanenburg2020image}
A.~Zwanenburg et~al.,
\newblock The Image Biomarker Standardization Initiative: standardized
  quantitative radiomics for high-throughput image-based phenotyping,
\newblock Radiology, 191145 (2020).

\bibitem{erasmus2004short}
L.~Erasmus, D.~Hurter, M.~Naud{\'e}, H.~Kritzinger, and S.~Acho,
\newblock A short overview of MRI artefacts,
\newblock SA Journal of Radiology {\bf 8} (2004).

\bibitem{Bushberg}
J.~T. Bushberg, J.~A. Seibert, E.~M. Leidholdt, and J.~M. Boone,
\newblock {\em {The essential physics of medical imaging}},
\newblock Lippincott Williams \& Wilkins, 2011.

\bibitem{um2019impact}
H.~Um, F.~Tixier, D.~Bermudez, J.~O. Deasy, R.~J. Young, and H.~Veeraraghavan,
\newblock Impact of image preprocessing on the scanner dependence of
  multi-parametric MRI radiomic features and covariate shift in
  multi-institutional glioblastoma datasets,
\newblock Physics in Medicine \& Biology {\bf 64}, 165011 (2019).

\bibitem{schwier2019repeatability}
M.~Schwier, J.~van Griethuysen, M.~G. Vangel, S.~Pieper, S.~Peled, C.~Tempany,
  H.~J. Aerts, R.~Kikinis, F.~M. Fennessy, and A.~Fedorov,
\newblock Repeatability of multiparametric prostate MRI radiomics features,
\newblock Scientific reports {\bf 9}, 1--16 (2019).

\bibitem{gardner1995detection}
E.~A. Gardner, J.~H. Ellis, R.~J. Hyde, A.~M. Aisen, D.~J. Quint, and P.~L.
  Carson,
\newblock Detection of degradation of magnetic resonance (MR) images:
  comparison of an automated MR image-quality analysis system with trained
  human observers,
\newblock Academic Radiology {\bf 2}, 277--281 (1995).

\bibitem{esteban2017mriqc}
O.~Esteban, D.~Birman, M.~Schaer, O.~O. Koyejo, R.~A. Poldrack, and K.~J.
  Gorgolewski,
\newblock MRIQC: Advancing the automatic prediction of image quality in MRI
  from unseen sites,
\newblock PloS one {\bf 12} (2017).

\bibitem{Ieremeiev2018}
O.~Ieremeiev, V.~Lukin, N.~Ponomarenko, and K.~Egiazarian,
\newblock {Robust linearized combined metrics of image visual quality},
\newblock Electronic Imaging {\bf 2018}, 260--1--260--6 (2018).

\bibitem{woodard2006no}
J.~P. Woodard and M.~P. Carley-Spencer,
\newblock No-reference image quality metrics for structural MRI,
\newblock Neuroinformatics {\bf 4}, 243--262 (2006).

\bibitem{VanEssen2012}
D.~C. {Van Essen} et~al.,
\newblock {The Human Connectome Project: A data acquisition perspective},
\newblock NeuroImage {\bf 62}, 2222--2231 (2012).

\bibitem{Mortamet2009}
B.~Mortamet, M.~A. Bernstein, C.~R. Jack, J.~L. Gunter, C.~Ward, P.~J. Britson,
  R.~Meuli, J.~P. Thiran, and G.~Krueger,
\newblock {Automatic quality assessment in structural brain magnetic resonance
  imaging},
\newblock Magnetic Resonance in Medicine {\bf 62}, 365--372 (2009).

\bibitem{Pizarro2016}
R.~A. Pizarro, X.~Cheng, A.~Barnett, H.~Lemaitre, B.~A. Verchinski, A.~L.
  Goldman, E.~Xiao, Q.~Luo, K.~F. Berman, J.~H. Callicott, D.~R. Weinberger,
  and V.~S. Mattay,
\newblock {Automated quality assessment of structural magnetic resonance brain
  images based on a supervised machine learning algorithm},
\newblock Frontiers in Neuroinformatics {\bf 10} (2016).

\bibitem{esteban2019crowdsourced}
O.~Esteban, R.~W. Blair, D.~M. Nielson, J.~C. Varada, S.~Marrett, A.~G. Thomas,
  R.~A. Poldrack, and K.~J. Gorgolewski,
\newblock Crowdsourced MRI quality metrics and expert quality annotations for
  training of humans and machines,
\newblock Scientific data {\bf 6}, 1--7 (2019).

\bibitem{KLAPWIJK2019116}
E.~T. Klapwijk, F.~van~de Kamp, M.~van~der Meulen, S.~Peters, and L.~M.
  Wierenga,
\newblock Qoala-T: A supervised-learning tool for quality control of FreeSurfer
  segmented MRI data,
\newblock NeuroImage {\bf 189}, 116 -- 129 (2019).

\bibitem{vogelbacher2019lab}
C.~Vogelbacher, M.~H. Bopp, V.~Schuster, P.~Herholz, A.~Jansen, and J.~Sommer,
\newblock LAB--QA2GO: A free, easy-to-use toolbox for the quality assessment of
  magnetic resonance imaging data,
\newblock Frontiers in neuroscience {\bf 13}, 688 (2019).

\bibitem{janowczyk2019histoqc}
A.~Janowczyk, R.~Zuo, H.~Gilmore, M.~Feldman, and A.~Madabhushi,
\newblock HistoQC: an open-source quality control tool for digital pathology
  slides,
\newblock JCO clinical cancer informatics {\bf 3}, 1--7 (2019).

\bibitem{pypi}
P.~S. Foundation,
\newblock The Python Package Index (PyPI),
\newblock \href{https://pypi.org/}{https://pypi.org/}.

\bibitem{FENG2017186}
{\sethlcolor{reviewer1}\hl{Y.~Feng, H.~Zhao, X.~Li, X.~Zhang, and H.~Li,
\newblock A multi-scale 3D Otsu thresholding algorithm for medical image
  segmentation,
\newblock Digital Signal Processing {\bf 60}, 186 -- 199 (2017).}}

\bibitem{TORRENTSBARRENA2019263}
{\sethlcolor{reviewer1}\hl{S.~Navdeep, K.~Lakhwinder, and S.~Kuldeep,
\newblock Histogram equalization techniques for enhancement of low radiance retinal images for early detection of diabetic retinopathy,
\newblock Engineering Science and Technology {\bf 22}, 736 -- 745 (2019).}}

\bibitem{scikit-image}
S.~van~der Walt, J.~L. {S}ch\"onberger, J.~{Nunez-Iglesias}, F.~{B}oulogne,
  J.~D. {W}arner, N.~{Y}ager, E.~{G}ouillart, T.~{Y}u, and the scikit-image
  contributors,
\newblock scikit-image: image processing in {P}ython,
\newblock PeerJ {\bf 2}, e453 (2014).


\bibitem{HUI201897}
{\sethlcolor{reviewer1}\hl{S.~Hui, T.~Zhang, L.~Shi, D.~Wang, C.~Ip, and W.~Chu,
\newblock Automated segmentation of abdominal subcutaneous adipose tissue and visceral adipose tissue in obese adolescent in MRI,
\newblock Magnetic Resonance Imaging {\bf 45}, 97 -- 104 (2018).}}

\bibitem{BINSAEEDAN202068}
{\sethlcolor{reviewer1}\hl{B.~Mnahi, A.~Faisal, F.~Faisal, A.~Khalefa, P.~Nadeem, and S.~Ghosh,
\newblock Check the chest: review of chest findings on abdominal MRI,
\newblock Clinical Imaging {\bf 59}, 68 -- 77 (2020).}}

  
\bibitem{7994618}
{\sethlcolor{reviewer1}\hl{S. Z. K. Sajib, N. Katoch, H. J. Kim, O. I. Kwon, and E. J. Woo,
\newblock Software Toolbox for Low-Frequency Conductivity and Current Density Imaging Using MRI,
\newblock IEEE Transactions on Biomedical Engineering {\bf 64}, 2505 -- 2514 (2017).}}


\bibitem{chang2015reference}
S.-J. Chang, S.~Li, A.~Andreasen, X.-Z. Sha, and X.-Y. Zhai,
\newblock A reference-free method for brightness compensation and contrast
  enhancement of micrographs of serial sections,
\newblock PloS one {\bf 10} (2015).

\bibitem{dietrich2007measurement}
O.~Dietrich, J.~G. Raya, S.~B. Reeder, M.~F. Reiser, and S.~O. Schoenberg,
\newblock Measurement of signal-to-noise ratios in MR images: influence of
  multichannel coils, parallel imaging, and reconstruction filters,
\newblock Journal of Magnetic Resonance Imaging: An Official Journal of the
  International Society for Magnetic Resonance in Medicine {\bf 26}, 375--385
  (2007).

\bibitem{edsall2003parallel}
R.~M. Edsall,
\newblock The parallel coordinate plot in action: design and use for geographic
  visualization,
\newblock Computational Statistics \& Data Analysis {\bf 43}, 605--619 (2003).

\bibitem{maaten2008visualizing}
L.~v.~d. Maaten and G.~Hinton,
\newblock Visualizing data using t-SNE,
\newblock Journal of machine learning research {\bf 9}, 2579--2605 (2008).

\bibitem{mcinnes2018umap}
L.~McInnes, J.~Healy, and J.~Melville,
\newblock Umap: Uniform manifold approximation and projection for dimension
  reduction,
\newblock arXiv preprint arXiv:1802.03426  (2018).

\bibitem{bakas2017advancing}
S.~Bakas, H.~Akbari, A.~Sotiras, M.~Bilello, M.~Rozycki, J.~S. Kirby, J.~B.
  Freymann, K.~Farahani, and C.~Davatzikos,
\newblock Advancing the cancer genome atlas glioma MRI collections with expert
  segmentation labels and radiomic features,
\newblock Scientific data {\bf 4}, 170117 (2017).

\bibitem{hui2010fast}
C.~Hui, Y.~X. Zhou, and P.~Narayana,
\newblock Fast algorithm for calculation of inhomogeneity gradient in magnetic
  resonance imaging data,
\newblock Journal of Magnetic Resonance Imaging {\bf 32}, 1197--1208 (2010).

\bibitem{antunes2020}
J.~T. Antunes, A.~Ofshteyn, K.~Bera, E.~Y. Wang, J.~T. Brady, J.~E. Willis,
  K.~A. Friedman, E.~L. Marderstein, M.~F. Kalady, S.~L. Stein, A.~S. Purysko,
  R.~Paspulati, J.~Gollamudi, A.~Madabhushi, and S.~E. Viswanath,
\newblock Radiomic Features of Primary Rectal Cancers on Baseline T2-Weighted
  MRI Are Associated With Pathologic Complete Response to Neoadjuvant
  Chemoradiation: A Multisite Study,
\newblock Journal of Magnetic Resonance Imaging {\bf n/a}.


\bibitem{636793}
A.~{Inselberg},
\newblock Multidimensional detective,
\newblock in {\em Proceedings of VIZ '97: Visualization Conference, Information
  Visualization Symposium and Parallel Rendering Symposium}, pages 100--107,
  1997.

\bibitem{ding2019supervised}
Y.~Ding, S.~Suffren, P.~Bellec, and G.~A. Lodygensky,
\newblock Supervised machine learning quality control for magnetic resonance
  artifacts in neonatal data sets,
\newblock Human brain mapping {\bf 40}, 1290--1297 (2019).

\bibitem{chirra2019multisite}
P.~Chirra, P.~Leo, M.~Yim, B.~N. Bloch, A.~R. Rastinehad, A.~Purysko, M.~Rosen,
  A.~Madabhushi, and S.~E. Viswanath,
\newblock Multisite evaluation of radiomic feature reproducibility and
  discriminability for identifying peripheral zone prostate tumors on MRI,
\newblock Journal of Medical Imaging {\bf 6}, 024502 (2019).

\bibitem{moradmand2019impact}
H.~Moradmand, S.~M.~R. Aghamir, and R.~Ghaderi,
\newblock Impact of image preprocessing methods on reproducibility of radiomic
  features in multimodal magnetic resonance imaging in glioblastoma,
\newblock Journal of Applied Clinical Medical Physics  (2019).

\bibitem{osadebey2018standardized}
M.~E. Osadebey et~al.,
\newblock Standardized quality metric system for structural brain magnetic
  resonance images in multi-center neuroimaging study,
\newblock BMC medical imaging {\bf 18}, 31 (2018).

\bibitem{doi:10.1002/mp.13245}
Y.~Wang, Y.~Zhang, W.~Xuan, E.~Kao, P.~Cao, B.~Tian, K.~Ordovas, D.~Saloner,
  and J.~Liu,
\newblock Fully automatic segmentation of 4D MRI for cardiac functional
  measurements,
\newblock Medical Physics {\bf 46}, 180--189 (2019).

\bibitem{http://bigwww.epfl.ch/publications/sage0303.html}
D.~Sage and M.~Unser,
\newblock Teaching Image-Processing Programming in {J}ava,
\newblock {IEEE} Signal Processing Magazine {\bf 20}, 43--52 (2003).

\bibitem{shehzad2015preprocessed}
Z.~Shehzad, S.~Giavasis, Q.~Li, Y.~Benhajali, C.~Yan, Z.~Yang, M.~Milham,
  P.~Bellec, and C.~Craddock,
\newblock The Preprocessed Connectomes Project Quality Assessment Protocol—a
  resource for measuring the quality of MRI data,
\newblock Frontiers in neuroscience {\bf 47} (2015).

\bibitem{monti2003consensus}
S.~Monti, P.~Tamayo, J.~Mesirov, and T.~Golub,
\newblock Consensus clustering: a resampling-based method for class discovery
  and visualization of gene expression microarray data,
\newblock Machine learning {\bf 52}, 91--118 (2003).

\bibitem{wilkerson2010consensusclusterplus}
M.~D. Wilkerson and D.~N. Hayes,
\newblock ConsensusClusterPlus: a class discovery tool with confidence
  assessments and item tracking,
\newblock Bioinformatics {\bf 26}, 1572--1573 (2010).

\bibitem{Beigclincanres.2556.2019}
{\sethlcolor{reviewer1}\hl{N.~Beig, K.~Bera, P.~Prasanna, J.~Antunes, R.~Correa, S.~Singh,
  A.~Saeed~Bamashmos, M.~Ismail, N.~Braman, R.~Verma, V.~B. Hill,
  V.~Statsevych, M.~S. Ahluwalia, V.~Varadan, A.~Madabhushi, and P.~Tiwari,
\newblock Radiogenomic-based survival risk stratification of tumor habitat on
  Gd-T1w MRI is associated with biological processes in Glioblastoma,
\newblock Clinical Cancer Research {\bf 26} 1866--1876  (2020).}}

\bibitem{rohlfing2010sri24}
{\sethlcolor{reviewer1}\hl{T.~Rohlfing, N.~M. Zahr, E.~V. Sullivan, and A.~Pfefferbaum,
\newblock The SRI24 multichannel atlas of normal adult human brain structure,
\newblock Human brain mapping {\bf 31}, 798--819 (2010).}}

\bibitem{KESHAVAN2018365}
{\sethlcolor{reviewer2}\hl{A.~Keshavan, E.~Datta, I.~McDonough, C.~Madan, K.~Jordan, and R.~Henry,
\newblock Mindcontrol: A web application for brain segmentation quality control,
\newblock NeuroImage {\bf 170}, 365--372 (2018).}}

\bibitem{pradeep_reddy_raamana_2018_1211365}
{\sethlcolor{reviewer2}\hl{P.R. Raamana,
\newblock Mindcontrol: VisualQC: Assistive tools for easy and rigorous quality control of neuroimaging data, (2018).}}

\bibitem{davatzikos2018cancer}
{\sethlcolor{reviewer1}\hl{C.~Davatzikos et~al.,
\newblock Cancer imaging phenomics toolkit: quantitative imaging analytics for
  precision diagnostics and predictive modeling of clinical outcome,
\newblock Journal of medical imaging {\bf 5}, 011018 (2018).}}

\bibitem{smith1997susan}
{\sethlcolor{reviewer1}\hl{S.~M. Smith and J.~M. Brady,
\newblock SUSAN: a new approach to low level image processing,
\newblock International journal of computer vision {\bf 23}, 45--78 (1997).}}

\bibitem{tustison2010n4itk}
{\sethlcolor{reviewer1}\hl{N.~J. Tustison, B.~B. Avants, P.~A. Cook, Y.~Zheng, A.~Egan, P.~A. Yushkevich,
  and J.~C. Gee,
\newblock N4ITK: improved N3 bias correction,
\newblock IEEE transactions on medical imaging {\bf 29}, 1310--1320 (2010).}}

\bibitem{Sunoqrot2020}
{\sethlcolor{reviewer1}\hl{M.~R.~S. Sunoqrot, G.~A. Nketiah, K.~M. Selnaes, T.~F Bathen, 
  and M. Elschot,
\newblock Automated reference tissue normalization of T2-weighted MR images of the prostate using object recognition,
\newblock Magnetic Resonance Materials in Physics, Biology and Medicine 1--13 (2020).}}

\bibitem{nyul2000stdn}
{\sethlcolor{reviewer1}\hl{L.~G. Nyul, J.~K. Udupa, and X. Zhang,
\newblock New variants of a method of MRI scale standardization,
\newblock IEEE transactions on medical imaging {\bf 19}, 143--150 (2000).}}

\bibitem{Scalco2020}
{\sethlcolor{reviewer1}\hl{E. Scalco,  A. Belfatto, A. Mastropietro, T. Rancati, B. Avuzzi, A. Messina, R. Valdagni, and G. Rizzo,
\newblock T2w-MRI signal normalization affects radiomics features reproducibility,
\newblock Medical Physics {\bf 47}, 1680--1691  (2020).}}

\bibitem{delis2017moving}
{\sethlcolor{reviewer2}\hl{H. Delis,  K. Christaki, B. Healy, G. Loreti, G.~L. Poli, P. Toroi, and A. Meghzifene,
\newblock Moving beyond quality control in diagnostic radiology and the role of the clinically qualified medical physicist,
\newblock Physica Medica {\bf 41}, 104--108  (2017).}}

\bibitem{Silvia2008}
{\sethlcolor{reviewer2}\hl{S. Ondategui-Parra,
\newblock Quality Management in Radiology: Defining the Parameters,
\newblock Health Management {\bf 8(4)}, (2008).}}

\end{thebibliography}

\end{document}